\documentclass[11pt]{article}
\usepackage{times,epsfig}

\newtheorem{theorem}{Theorem}
\newtheorem{lemma}{Lemma}
\newtheorem{define}{Definition}
\newtheorem{corollary}{Corollary}

\newtheorem{proposition}{Proposition}

\title{Asymptotic Behavior of Error Exponents in the Wideband Regime
\thanks{Research supported by DARPA grant F30602-00-2-0542, AFOSR URI
  grant F49620-01-1-0365 and NSF ITR grant 00-85929}
}

\author{Xinzhou Wu and R. Srikant\\
Coordinated Science Lab\\
University of Illinois at Urbana-Champaign\\
xwu@uiuc.edu, rsrikant@uiuc.edu}

\begin{document}
\maketitle

\begin{abstract}
In this paper, we complement Verd\'{u}'s work on spectral
efficiency in the wideband regime by investigating the fundamental
tradeoff between rate and bandwidth when a constraint is imposed
on the error exponent. Specifically, we consider both AWGN and
Rayleigh-fading channels. For the AWGN channel model, the optimal
values of $R_z(0)$ and $\dot{R_z}(0)$ are calculated, where
$R_z(1/B)$ is the maximum rate at which information can be
transmitted over a channel with bandwidth $B/2$ when the
error-exponent is constrained to be greater than or equal to $z.$
Based on this calculation, we say that a sequence of input
distributions is near optimal if both $R_z(0)$ and $\dot{R_z}(0)$
are achieved. We show that QPSK, a widely-used signaling scheme,
is near-optimal within a large class of input distributions for
the AWGN channel. Similar results are also established for a
fading channel where full CSI is available at the receiver.
\end{abstract}


\section{Introduction} \label{sec: intro}

Communications in the wideband regime with limited power has
attracted much attention recently. An important characteristic of
such communication systems is that they operate at relatively low
spectral efficiency (bits per second per Hz) and energy per bit.
The advantages of communication over large bandwidth are
many-fold: power savings, higher data rates, more diversity to
combat frequency-selective fading, etc. Thus, it is important to
understand the ultimate limits of communications in this regime
from an information-theoretic point of view, and develop
guidelines to design good signaling schemes.

Communications without a bandwidth limit, i.e., the available
bandwidth is infinite, is well understood. For the additive white
Gaussian noise (AWGN) channel, the capacity, measured in nats per
second, converges to the signal-to-noise ratio (SNR) $P/N_0$ of
the channel when the available bandwidth $B$ goes to infinity.
Here $P$ denotes the average power constraint at the input of the
channel and $N_0/2$ is the power-spectral density of the Gaussian
noise. Furthermore, a Gaussian signaling scheme is not mandatory
to achieve this limit. Nearly all signaling schemes are equally
good in the sense that the corresponding mutual information
converges to the same value in the infinite bandwidth limit. For
example, a simple on-off signaling scheme with low duty cycle is
capacity-achieving in the infinite bandwidth limit. In
\cite{mas76}, Massey showed that all mean zero signaling schemes
can achieve this limit.

\begin{figure}\label{fig: infinite bandwidth}
\center \epsfig{file=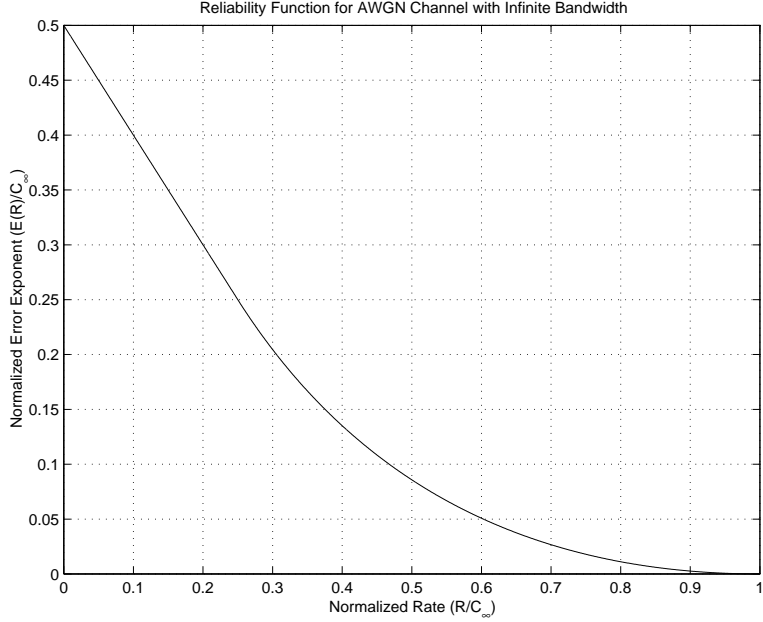,width=4.0in} \caption{The
reliability function for AWGN channel with infinite bandwidth}
\end{figure}

To establish a {\em strong coding theorem}, the {\em reliability
function} $E(R)$, as defined in \cite{gal68}, of the channel has
to be calculated for any coding rate $R$. Generally, the
reliability function of a channel is difficult to compute and is
known for all rates only for a few channels. Infinite-bandwidth
AWGN channel is one of these channels and its reliability function
has the following form\cite{wozjac65,gal68}
\begin{equation}
E(R)=\left\{
\begin{array}{ll}
\frac{C_{\infty}}{2}-R&0\le R\le\frac{C_{\infty}}{4};\\
(\sqrt{C_{\infty}}-\sqrt{R})^2&\frac{C_{\infty}}{4}\le R\le C_{\infty},\\
\end{array}
\right.\label{eq: infinite bandwidth E}
\end{equation}
where $C_{\infty}=P/N_0$ denotes the infinite-bandwidth capacity,
as shown in Figure~\ref{fig: infinite bandwidth}. We will show
that when the bandwidth is infinite, a large set of input
distributions can be shown to achieve the optimal error-exponent
curve. We will refer to such distributions as being {\em
first-order optimal}.

Naturally, the results in the infinite bandwidth regime can be
considered as guidelines for designing signaling schemes in the
wideband regime as well. However, in the wideband regime (when the
available bandwidth is large, but finite), the result based on the
infinite bandwidth calculations can be quite misleading. In
\cite{ver02}, Verd\'{u} points out that to understand the
performance limit in the wideband regime, two quantities need to
be studied: the minimum energy per information bit
$({\frac{E_b}{N_0}}_{min})$ required to sustain reliable
communication, and the slope of spectral efficiency (bits/s/Hz) at
the point ${\frac{E_b}{N_0}}_{min}.$ If we treat $C(\cdot)$ as a
function of $b=1/B$, it is easy to see that studying these two
quantities is equivalent to studying the optimal values of the
following two quantities: infinite-bandwidth capacity $C(0)$ and
the first-order derivative of capacity with respect to $b$,
$\dot{C}(0).$ In other words, we need to study both the
infinite-bandwidth capacity, and the rate at which this capacity
is reached. In \cite{ver02}, it is shown that, while many
signaling schemes achieve $C(0),$ only some of these reach the
capacity at the fastest possible rate given by $\dot{C}(0).$ We
will refer to signaling schemes that achieve both $C(0)$ and
$\dot{C}(0)$ as {\em near-optimal} input distributions in the
wideband regime. Further, although $C(0)$ always has the same
value for non-fading or fading channels with different CSI,
$\dot{C}(0)$ is determined by the CSI and can be very different
for different channels.

This paper complements Verd\'{u}'s work and considers the
relationship between probability of decoding error (represented by
the reliability function), coding rate, and bandwidth for both
AWGN channels and multi-path fading channels. Specifically, we
study the maximum rate at which information can be transmitted
over a channel, as a function of the available bandwidth, under a
certain constraint on the reliability function. For AWGN channels,
instead of characterizing the capacity $C$ as a function of
$b=1/B$ as in \cite{ver02}, we are interested in characterizing
$R_z$ as a function of $b,$ where $R_z$ is the maximum rate such
that $E(R_z)\ge z$ and $E(R)$ is the reliability function of the
channel. In the infinite bandwidth regime, we characterize the
optimal rate $R_z(0)$ with respect to a certain error-exponent
constraint and study the conditions under which a signaling scheme
can achieve this optimal rate. In the wideband regime, both
$R_z(0)$ and $\dot{R_z}(0)$ need to be considered. A signaling
scheme which can achieve both $R_z(0)$ and $\dot{R_z}(0)$ is said
to be {\em second-order optimal} or {\em near optimal} with
respect to an error-exponent constraint $z.$


For fading channels, we use a doubly-block fading model where the
available bandwidth spans multiple coherence bandwidth. If we let
$W_c$ denote the coherence bandwidth, the total bandwidth of the
channel is then assumed to $BW_c$ for some $B\geq 1.$  Either a
large $B$ or a large $W_c$ can lead to a large total bandwidth
$BW_c.$ However, these two regimes (the large $B$ regime and the
large $W_c$ regime) can have very different channel behavior.
Suppose we consider a wireless system with a total bandwidth of
$10$ MHz and if the delay spread is of the order of $1$ $\mu$sec.,
then $W_c$ would be of the order of $1$ MHz and thus, $B$ is of
the order of $10.$ In this paper, we focus on such a system where
the coherence bandwidth $W_c$ is large and further, we assume a
coherent channel model. By defining $R_z$ to be a function of
$1/W_c,$ we calculate $R_z(0)$ and $\dot{R}_z(0).$ Similar to the
AWGN case, for this channel model, we will show that QPSK can
achieve both  $R_z(0)$ and $\dot{R}_z(0)$ and is thus
near-optimal. In the other case where $B$ is large, it may not be
appropriate to assume any form of channel side information (CSI)
and thus a non-coherent channel model is more suitable. We refer
the readers to \cite{wusri04} for first-order asymptotic results
for MIMO channels in this regime.

This paper is organized as follows. In section~\ref{sec: channel
model}, we will specify the channel models and formulate the
problem that we wish to study. In section~\ref{sec: main results},
we will show the main results for both AWGN channels and multipath
fading channels. The proofs will be presented in section~\ref{sec:
AWGN proof} and section~\ref{sec: fading proof}. Section~\ref{sec:
conclusions} contains concluding remarks and discussions.

\section{Channel models and problem formulation}\label{sec: channel model}

In this section, we will describe the channel models we use to
study the behavior of both the AWGN channel and the multipath
fading channel in the wideband regime. Further, we will formulate
rigorously the problems we want to solve in this paper.

\subsection{AWGN channels}

We first consider a bandlimited AWGN channel with available
bandwidth $B/2:$
\begin{equation}
y(t)=x(t)+w(t),\label{eq: waveform}
\end{equation}
where $w(t)$ is a complex symmetric Gaussian random process. We
assume that we have an input power constraint $P$ for the channel
(\ref{eq: waveform}). For notational convenience, we assume the
noise power density $N_0/2=1/2.$ Thus, the average power $P$ also
indicates the average SNR of the channel. We now sample the
channel at sampling rate $1/B,$  and represent it as a
discrete-time memoryless scalar channel as follows:
\begin{equation}
y=x+w, \enskip \label{eq: scalar channel}
\end{equation}
where $w$ is a complex symmetric Gaussian random variable with
variance $1,$ i.e., $w\in {\mathcal CN}(0,1).$ The power
constraint for this discrete-time channel is
\begin{equation}
E\left(|x|^2\right) \le \frac{P}{B}. \label{eq: power constraint}
\end{equation}
We want to study the asymptotic behavior of the communication rate
$R$ (nats per second) in terms of the available bandwidth $B$
under this power constraint and an error exponent constraint,
which is described below.

Let $P_e(N,R,P,B)$ be the minimum probability of decoding error
for any block code with codeword length $N$ seconds (or
equivalently, $NB$ symbols) and coding rate $R.$ The error
exponent at communication rate $R$ (also called {\em reliability
function}) of this channel is defined as
\begin{equation}
E(R,P,B)=\lim_{N\rightarrow \infty} -\frac{\ln
P_e(N,R,P,B)}{N}.\label{eq: def E}
\end{equation}

We desire a lower bound for $E(R,P,B)$ and denote it by $Pz.$
(Without loss of generality, we scale  the desired minimum value
for the error exponent by $P$ for mathematical convenience.) Let
$R_z(b)$ denote the maximum possible rate at which communication
is possible given this desired error exponent when the available
bandwidth is $B=1/b.$ Since $E(P,R,B)$ is a decreasing function of
$R,$ $R_z(b)$ is the solution to the equation
\begin{equation}\label{eq: main problem}
E(P,R,1/b)=Pz.
\end{equation}
Our goals for AWGN channels are two-folds:
\begin{itemize}
\item[1.] Calculate $R_z(0)$ and $\dot{R_z}(0).$ \item[2.]
Characterize the properties of {\em first-order optimal} signaling
schemes, i.e., those that achieve $R_z(0).$ More importantly, find
{\em near-optimal} or {\em second-order optimal} signaling schemes
in the wideband regime such that both $R_z(0)$ and $\dot{R}_z(0)$
can be achieved.
\end{itemize}

In the rest of the paper, we drop the subscript and simply refer
to $R_z$ as $R.$ From the context, it should be clear that $R$ is
a function of $z.$

\subsection{Coherent fading channels}

In this section, we will explain the model we will use for a
multi-path fading channel and formulate the problem in the
wideband regime we want to solve for such channels.

To characterize a multi-path fading channel, we use a {\em
doubly-block} Rayleigh fading model. Specifically, we assume block
fading in both the time and frequency domains. Further, we assume
that we have a rich-scattering environment such that all the
fading gains are Gaussian distributed. This model can be
visualized as in Figure~\ref{fig: frequencytime}, where we divide
the time-frequency plane into blocks of duration $T_c$ and
bandwidth $W_c.$ We assume that the fading is fixed in each block
and independent from one block to another. In each block, we can
transmit $W_cT_c$ symbols, from the {\em dimensionality theorem}
\cite{wozjac65}. We let $D=W_cT_c$ and refer to $D$ as the {\em
coherence dimension} of the channel.

\begin{figure}
\center \epsfig{file=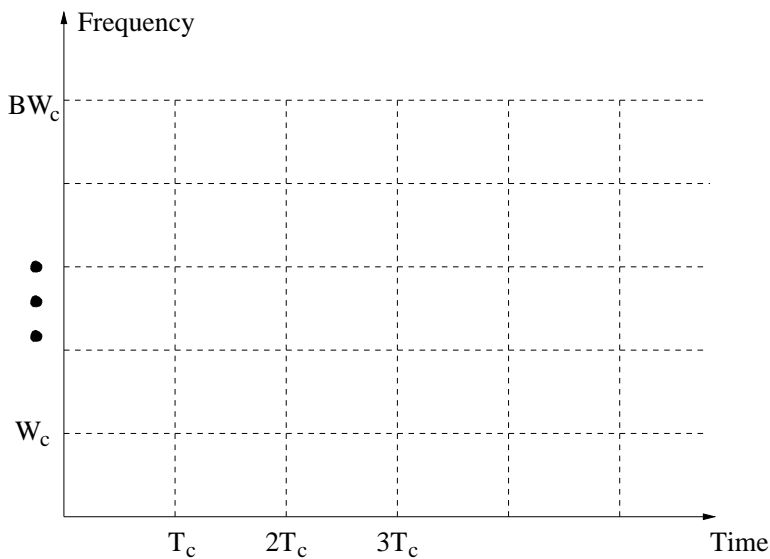,width=4.0in}
\caption{Doubly-block fading in time-frequency plane}\label{fig:
frequencytime}
\end{figure}

For this channel model, we can represent the channel by
\begin{equation}
{\bf y}_l=H_l{\bf x}_l+{\bf w}_l, \quad 1\le l\le B,\label{eq:
fading vector channel}
\end{equation}
where ${\bf x}_l, {\bf y}_l, {\bf w}_l\in {\mathcal{C}^{D}}.$ In
other words, we have $B$ parallel vector channels each with
dimension $D.$ Similar to the AWGN channel, we assume there is
power constraint $P$ (joule per second) for the fading channel,
i.e., we have the following constraint on the input of the channel
(\ref{eq: fading vector channel}):
\begin{equation}
\sum_{l=1}^B E[\|{\bf x}_l\|^2]\le PT_c.\label{eq: power
constraint B}
\end{equation}

The doubly-block fading model is a simple approximation of the
physical multipath fading channel. However, it retains most of the
important characteristics of channels in a fading environment. For
a derivation of such a model, we refer the interested reader to
\cite{tsevis04}. This model has been used in \cite{medtse00} to
achieve the lower bound for the optimal bandwidth where spreading
still increases non-coherent channel capacity. In \cite{hajsub02},
Hajek and Subramanian use this model to calculate the reliability
function and capacity for a non-coherent fading channel with a
small peak constraint on the input signals. However, this model is
simpler than the model used by M\'{e}dard and Gallager
\cite{medgal02}, which allows correlation in both time and
frequency blocks, or the model used Telatar and Tse
\cite{teltse00}, which allows correlation in frequency blocks.

In the wideband regime, we know the available bandwidth $BW_c>>1$
and the energy available per degree of freedom is small, i.e.,
$\frac{P}{BW_c}<<1.$ Obviously, a large bandwidth can be a result
of either a large $B$ or a large $W_c.$ However, $B$ and $W_c$
have different impacts on the channel performance and the
asymptotic results in $B$ and $W_c$ can be very different from
each other and can lead to different conclusions. In this paper,
we will focus on the case where $W_c$ is large. In this regime, we
have large degrees of freedom in each coherence block although the
energy per degree of freedom is small. Thus, we might still be
able to measure the channel accurately and therefore, we assume a
coherent fading channel model in this regime. However, to
accurately illustrate the coherence level of this channel model
from an error exponent point of view is still a research topic for
now. We refer the reader to \cite{zhemed03} for a discussion on
the relationship between coherence level and coherence length from
a capacity point of view.

The ergotic capacity of such channels under full receiver side CSI
is well known and is determined by the following expression
\begin{equation}
C=BW_cE_{H}[\ln(1+\frac{|H|^2P}{BW_c})] \quad \textsl{nats per
second}.\label{eq: fc}
\end{equation}
The reliability function $E(R,P,W_c)$ of this channel can be
defined as below
\begin{equation}
E(R,P,W_c)=\lim_{N\rightarrow \infty} -\frac{1}{T_c}\frac{\ln
P_e(N,R,P,W_c)}{N},\label{eq: fee}
\end{equation}
where $P_e(N,R,P,W_c)$ is the minimum probability of decoding
error for all block codes with codeword length $NT_c$ seconds and
coding rate $R$ (nats per second).

Let $R_z(1/W_c)$ denote the maximum possible rate at which
communication is possible given this desired error exponent
$E(R,P,W_c)\ge z.$ Our goal in studying this channel model in the
wideband regime is still two-fold: calculate both $R_z(0)$ and
$\dot{R}_z(0)$ and identify signaling schemes that can achieve
$R_z(0)$ and $\dot{R}_z(0).$

\section{Main results}\label{sec: main results}
In this section, we will present our main results for AWGN
channels and coherent fading channels in two separate sections
without proof. Due to the technical nature of the proofs, we will
present them in Section~\ref{sec: AWGN proof} and
Section~\ref{sec: fading proof}.

\subsection{AWGN channels}
We begin by first carefully describing the set of signaling
schemes that we will consider in this paper. Due to the
technicality in applying the sphere-packing bound (see
Appendix~\ref{sec: reliability function} for a short review), we
only consider input distributions with a finite alphabet.
Specifically, we restrict ourselves to input distributions in the
following set.

\begin{define}\label{def: Dp}
Define
$${\mathcal D}(p)=\{q(x): E[|x|^2]=p; \textsl{support of q(x)
is a finite set of discrete points in ${\mathcal C}$}\}.$$
\end{define}

We impose the following additional constraint on the signaling
schemes.
\begin{define}\label{def: Qp}
Define ${\mathcal Q}(p)$ as a subset of ${\mathcal D}(p)$, which
satisfies the following properties
\begin{equation}
{\mathcal Q}(p)=\left\{ q_p(x)\in {\mathcal D}(p): |x|_{max}\le
K_m p^{\alpha}. \right\}
\end{equation}
where $|{x}|_{max}$ denotes the largest norm among all symbols of
the input alphabet. $K_m$ and $\alpha$ are allowed to be any
positive constants which are independent of $p.$ \hfill $\diamond$
\end{define}

In other words, we constrain the input such that the
largest-magnitude symbol has to decrease as $B$ increases,
although it can decrease at an arbitrarily slow rate. As we will
show later, the choice of the parameters $K_m$ and $\alpha$ are
not relevant to the result. Thus, $K_m$ can be an arbitrary large
number and $\alpha$ can be an arbitrary small positive number, if
we want to make the constraint mild.

A signaling scheme is a sequence of input distributions,
parameterized by $B.$ For each $B,$ we can only choose an input
distribution from the set ${\mathcal Q}(P/B).$

\begin{define}\label{def: F(P)}
We define ${\mathcal F}(P)$ to be the set of signaling schemes,
which are parameterized by $B$ and satisfy
\begin{equation}
{\mathcal F}(P)=\left\{ \left\{q_B(x)\right\}:q_B(x)\in {\mathcal
Q}(P/B)\right\},
\end{equation}
where ${\mathcal Q}(P/B)$ is defined by Definition~\ref{def: Qp}.
\hfill $\diamond$
\end{define}

By choosing signaling schemes from ${\mathcal F}(P),$ we are
ruling out those {\em peaky} signaling schemes in which one of the
input symbols remains constant or goes to $\infty,$ while the
average power per degree of freeedom goes to $0.$

Under these constraints on the input distribution, we now specify
the reliability function $E(R,P,B)$ defined by (\ref{eq: def E})
for AWGN channels.

\begin{lemma}\label{lem: bounds}
Consider the discrete-time additive Gaussian channel (\ref{eq:
scalar channel}) with bandwidth $B/2$ and input signaling schemes
constrained by ${\mathcal F}(P).$ Then the reliability function
for this channel satisfies
\begin{equation}
E_r(R,P,B)\le E(R,P,B)\le E_{sp}(R,P,B),\label{eq: bounds}
\end{equation}
with
\begin{eqnarray}
&E_r(R,P,B)&=\sup_{0\le \rho\le 1} -\rho R+BE_o(P/B,\rho),\label{eq: rc}\\
&E_{sp}(R,P,B) &= \sup_{\rho\ge 0} -\rho R+BE_o(P/B,\rho),\nonumber\\
&E_o(P/B,\rho)&=\sup_{q\in{\mathcal Q}(P/B)}\sup_{\beta\ge 0}
-\ln\int\left(\int
q(x)e^{\beta\left(|x|^2-P/B\right)}f_w(y-x)^{\frac{1}{1+\rho}}
dx\right)^{1+\rho} dy, \label{eq: Enote0 0}
\end{eqnarray}
where $f_w(x)$ is the probability density function of a complex
Gaussian random variable ${\mathcal CN}(0,1).$
\end{lemma}
{\bf Proof:} This directly follows from the discussion on error
exponent in Appendix~\ref{sec: reliability function}. \hfill
$\diamond$

{\bf Remarks:} The most important fact here is that as we pointed
out in Appendix~\ref{sec: reliability function}, there exists a
{\em critical rate} $R_{crit},$ such that for $R\ge R_{crit},$ the
sphere packing bound and the random-coding bound coincide with
each other and thus the random-coding exponent (\ref{eq: rc}) with
(\ref{eq: Enote0 0}) actually is the true reliability function.
Based on this fact, if we only focus on this rate region, by
characterizing the asymptotic behavior of (\ref{eq: rc}) when $B$
is large, we get the asymptotic behavior of the reliability
function. In the following theorem, we obtain closed-form
expressions for $R(0)$ and $\dot{R}(0).$

\begin{theorem}\label{thm: non-fading}
Consider the discrete-time additive Gaussian channel (\ref{eq:
scalar channel}) with bandwidth $B/2$ and input signaling schemes
constrained by ${\mathcal F}(P).$ Let $R(1/B)$ be the maximum rate
at which information can be transmitted on this channel such that
the following error-exponent constraint is satisfied:
\begin{equation}
E(R,P,B)\ge Pz, \quad 0<z<\frac{1}{4}.\label{eq: error constraint}
\end{equation}
We have
\begin{equation}
R(0)=\lim_{B\rightarrow \infty} R(1/B) =
P(1-\sqrt{z})^2,\label{eq: R zero}
\end{equation}
and
\begin{equation}
\dot{R}(0)= -\frac{P^2 (1-\sqrt{z})^3}{2}.\label{eq: dotR}
\end{equation}
\hfill $\diamond$
\end{theorem}
{\bf Remarks:} The constraint on $z$ in (\ref{eq: error
constraint}) arises from the fact that the reliability function is
only determined for a certain range of $z.$ Outside this range,
the random-coding exponent is not necessarily tight. As we will
show later, $z=\frac{1}{4}$ is the error exponent for $R=R_{crit}$
in the infinite bandwidth limit. We now argue that for
$0<z<\frac{1}{4},$ when the bandwidth is sufficiently large, the
solution $R(1/B)$ to (\ref{eq: error constraint}) will exceed
$R_{crit}(1/B)$ and thus, the error exponent at $R(1/B)$ is equal
to the random-coding exponent. To be precise, we state this
argument in the following lemma and provide the proof in the
appendix. It follows from this lemma that we can represent the
reliability function by the random-coding exponent if we only
consider $z<\frac{1}{4}.$
\begin{lemma}\label{lem: quarter}
Let $R_r(1/B)$ be the solution to the random-coding exponent
constraint $E_r(R,P,B)=Pz,$ for a fixed $z\in(0,\frac{1}{4}).$ For
a fixed $z<\frac{1}{4},$ we must be able to find a $B_z<\infty,$
such that for all $B>B_z,$ $R(1/B)=R_r(1/B).$
\end{lemma}
{\bf Proof:} See Appendix~\ref{sec: quarter}. \hfill $\diamond$

It should be noted that the constraints on the input signaling are
not necessary to obtain the first-order result (\ref{eq: R zero}). 
In other words, introducing {\em peakiness} or allowing continuous
alphabet symbols in the input distributions will not improve the
error exponent in the infinite bandwidth limit for the AWGN
channel. These constraints only play a role in obtaining the
second-order terms in the expansion of $R_z(1/B)$ around $1/B=0.$

A main goal of our study of the wideband reliability function here
is to find good signaling schemes in the sense that they can
achieve $R(0)$ and $\dot{R}(0).$ To do that, we first define {\sl
first-order optimality} and {\sl near optimality} (or {\sl
second-order optimality}) formally of a signaling scheme in the
wideband regime, in a similar way as in \cite{ver02}.
\begin{define}
Consider a signaling scheme $\{q_B({\bf x})\}\in {\mathcal F}(P)$
parameterized by $B.$ Let $\tilde R(1/B)$ be the solution of
\begin{equation}
Pz=E(R,q_B,P,B) \label{eq: s00}
\end{equation}
where $E(R,q_B,P,B)$ is the reliability function of the channel
when the input distribution is fixed to be $q_B.$ This signaling
scheme is said to be {\sl first-order optimal} with respect to the
normalized error exponent $z$, if
$${\tilde R}(0)={R}(0).$$ \hfill $\diamond$
\end{define}

\begin{define}
A signaling scheme $\{q_B({\bf x})\}\in {\mathcal F}(P)$ is called
{\sl second-order optimal} or {\sl near optimal} with respect to
the normalized error exponent $z$ if
\begin{eqnarray}
&&{\tilde R}(0)=R(0);\\
&&\dot{\tilde R}(0)=\dot{R}(0),
\end{eqnarray}
where $\tilde R(1/B)$ is the solution to (\ref{eq: s00}). \hfill
$\diamond$
\end{define}

For AWGN channels, we obtain a sufficient condition for a
signaling scheme to be first-order optimal. Then, we study the
performance of two simple signaling schemes as in \cite{ver02}:
BPSK and QPSK. Specifically, when we say BPSK or QPSK, we mean the
following.  Let $p=P/B$ be the available power per degree of
freedom. For BPSK, we choose the input to be either $\sqrt{p}$ or
$-\sqrt{p}$ with equal probability; for QPSK, the input alphabet
consists of $\sqrt{\frac{p}{2}}(1+j)$, $\sqrt{\frac{p}{2}}(1-j)$,
$\sqrt{\frac{p}{2}}(-1+j)$, and $\sqrt{\frac{p}{2}}(-1-j)$, all
chosen with equal probability as well.

\begin{theorem}\label{thm: BPSK and QPSK}
For AWGN channels, all signaling schemes in ${\mathcal F}(P)$
which are symmetric around $0$ are first-order optimal for any
given $z\in(0,\frac{1}{4})$. Thus, both BPSK and QPSK are
first-order optimal; however, only QPSK is second-order
optimal.\hfill $\diamond$
\end{theorem}
{\bf Remarks:} From this theorem, we know that it does not take
much for a signaling scheme to be first-order optimal. This result
is consistent with the capacity result shown by Massey in
\cite{mas76}.

\begin{figure}
\center \epsfig{file=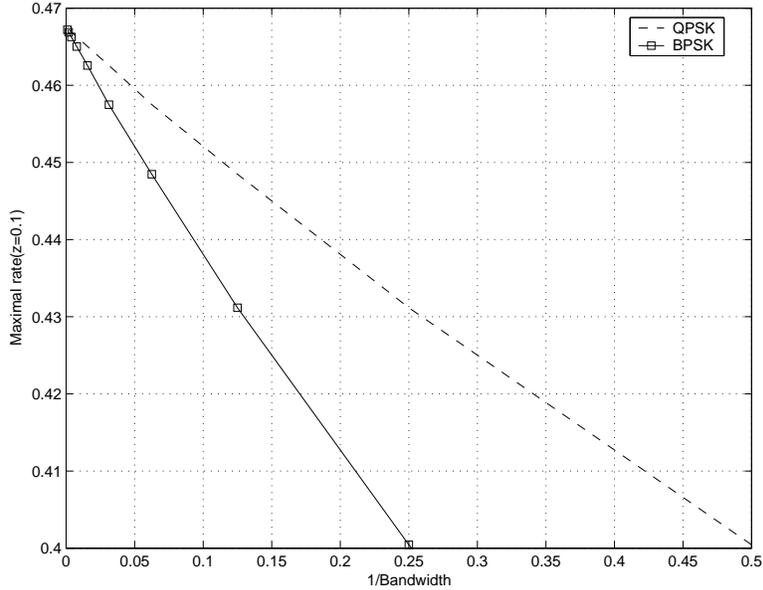,width=4.0in} \caption{The
maximal rate $R$ for BPSK and QPSK for a fixed normalized error
exponent $z=0.1.$}\label{fig: BPSK_QPSK1}
\end{figure}

To get a better feel for how differently BPSK and QPSK behave in
the wideband regime, we plot $R$ as a function of $1/B$ for both
BPSK and QPSK in Figure~\ref{fig: BPSK_QPSK1}. As shown in
Figure~\ref{fig: BPSK_QPSK1}, as $B\rightarrow \infty,$ both BPSK
and QPSK can achieve the optimal rate $R(0).$ However, only QPSK
can achieve $\dot{R}(0).$

\begin{figure}
\center \epsfig{file=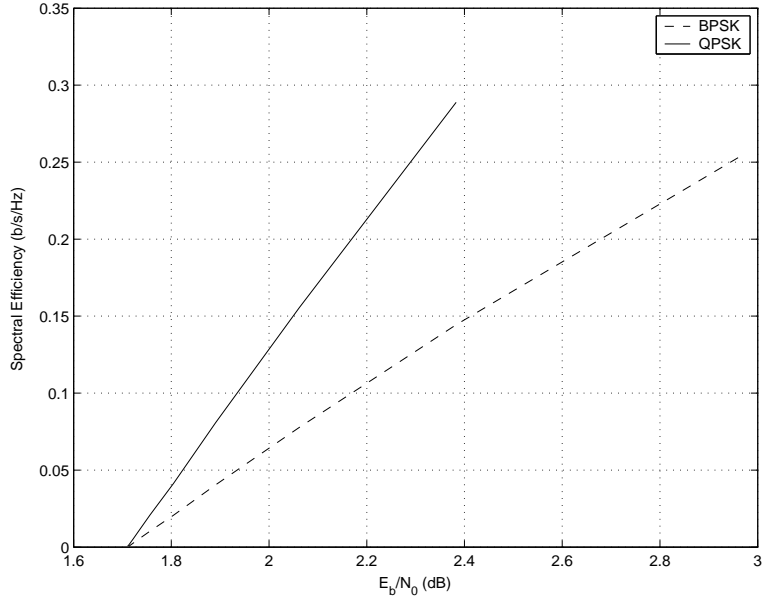,width=4.0in}
\caption{Spectral
  efficiencies achieved by QPSK and BPSK in the AWGN channel, when the
  error exponent is constrained by $z=0.1.$}\label{fig: SE}
\end{figure}

Another way to understand the difference between the performance
of BPSK and QPSK is to study the fundamental tradeoff between
spectral efficiency and energy per information bit ($E_b/N_0$), as
suggested in \cite{ver02}. We plot this tradeoff in
Figure~\ref{fig: SE}. From this figure, we can see that both BPSK
and QPSK can achieve the optimal ${\frac{E_b}{N_0}}_{min},$
however, only QPSK can achieve the optimal spectral efficiency
slope at the point ${\frac{E_b}{N_0}}_{min}.$

As compared to Figure 2 in \cite{ver02}, the major difference here
is that ${\frac{E_b}{N_0}}_{min}$ in Figure~\ref{fig: SE} is
around $3.3dB$ higher, since we have a more stringent constraint
than just reliable communications, as considered in \cite{ver02}.
${\frac{E_b}{N_0}}_{min}$ here denotes the minimal energy per
information bit such that the probability of error has to decay
faster than $e^{-Nz}$ as the codeword length $N$ increases.

\subsection{Coherent fading channels}

Next, we consider coherent fading channels. As in the case of the
AWGN channel, we first describe our assumptions on the input
signaling schemes.

\begin{define}\label{def: QBD(P)}
Define $\mathcal{Q}^B_{W_c}(P)$ to be the set of joint input
distributions on ${\bf X}=({\bf x}_1,{\bf x}_2,\cdots,{\bf x}_B),$
where $\{{\bf x}_l,\enskip l=1,2,\cdots,B\}$ are vectors with
dimension $D=W_cT_c,$ which satisfy the following
\begin{itemize}
\item[1.] the average power constraint (\ref{eq: power constraint
B}) is satisfied; \item[2.] the distribution has a discrete
alphabet, consisting of finite number of symbols; \item[3.] each
symbol can be chosen from a given set ${\mathcal S}^B_{W_c}.$ The
set of symbols ${\mathcal S}^B_{W_c}$ is defined as follows:
\begin{equation}
{\mathcal S}^B_{W_c}=\{{\bf X}=\{{\bf x}_1,{\bf x}_2,\cdots,{\bf
x}_B\}: {\bf x}_l\in C^D; \max_{d=1,2,\cdots D} |x_{ld}|\le K_m
W_c^{-\alpha} \enskip \forall l=1,2,\cdots,B\},
\end{equation}
where $K_m$ and $\alpha$ are allowed to be any positive constants
independent of $W_c.$ \hfill $\diamond$
\end{itemize}
\end{define}
The signaling schemes of interest to us are defined as follows.
\begin{define}\label{def: FBD(P)}
We define ${\mathcal F}^B_{W_c}(P)$ to be the set of signaling
schemes, which are parameterized by $W_c$ and satisfy
\begin{equation}
{\mathcal F}^B_{W_c}(P)=\left\{ \left\{q_{W_c}({\bf
X})\right\}:q_{W_c}({\bf X})\in {\mathcal
Q}^B_{W_c}(P)\right\},\label{eq: signal constraint B}
\end{equation}
where ${\mathcal Q}^B_{W_c}(P)$ was defined in
Definition~\ref{def: QBD(P)}. \hfill $\diamond$
\end{define}

The reliability function for our discrete-time channel model
(\ref{eq: fading vector channel}) with signaling schemes
constrained by ${\mathcal F}_{W_c}^B(P)$ can be computed according
to the following lemma.

\begin{lemma}\label{lem: fading bounds}
Consider the coherent fading channel model (\ref{eq: fading vector
channel}) with $H$ known at the receiver. Assume that the input
distribution satisfies the average power constraint (\ref{eq:
power constraint}) and the constraint in in ${\mathcal
F}_{W_c}^B(P).$ The reliability function $E(R,P,W_c)$ satisfies
$$
E_r(R,P,W_c)\le E(R,P,W_c)\le E_{sp}(R,P,W_c),
$$
with
\begin{eqnarray}
E_r(R,P,W_c)&=&\sup_{0\le \rho \le 1} -\rho R+E_o(P,\rho,W_c),\nonumber\\
E_{sp}(R,P,W_c)&=&\sup_{\rho \ge 0} -\rho R+E_o(P,\rho,D),\nonumber\\
E_o(P,\rho,W_c)&=& \sup_{q\in{\mathcal F}_{W_c}^B(P)}
\sup_{\beta\ge 0} -\frac{1}{T_c} \ln E_{H} \int\left(\int q({\bf
X})e^{ \beta(\|{\bf X}\|^2 -PT_c)} f({\bf Y}|{\bf X},{\bf
H})^{\frac{1}{1+\rho}} d{\bf X}\right)^{1+\rho} d{\bf Y}.
\label{eq: Enote}
\end{eqnarray}
\end{lemma}
{\bf Proof:} We can apply Theorem~\ref{thm: random coding} and
Theorem~\ref{thm: sphere packing} from Appendix~\ref{sec:
reliability function} 
here to this channel model by viewing the
channel as a memoryless channel with output $\hat{\bf Y}=\{{\bf
Y},{\bf H}\}.$ The fraction of $\frac{1}{T_c}$ in (\ref{eq:
Enote}) is to balance the scaling since the rate $R$ here is
defined to be nats per second. \hfill $\diamond$

The constraint on the error exponent is
\begin{equation}
E(R,P,W_c)\ge z,\label{eq: B problem}
\end{equation}
and we need to solve for $R(0)$ and $\dot{R}(0)$ where $R$ is a
function for $\frac{1}{W_c}$ for a fixed $B.$ We have the
following theorem.
\begin{theorem}\label{thm: fading D B}
Consider a coherent Rayleigh-fading vector channel (\ref{eq:
fading vector channel}) with the input signaling constrained by
${\mathcal F}^B_{W_c}(P).$ Let $R(1/W_c)$ be the maximum rate at
which information can be transmitted on this channel such that the
following error-exponent constraint is satisfied:
\begin{equation}
E(R,P,W_c)\ge z, \quad 0<z<z^*,\label{eq: fec}
\end{equation}
where $z^*$ is defined as follows
\begin{equation}
z^*=\frac{B}{T_c}\ln(1+\frac{PT_c}{2B})-\frac{P}{4+2PT_c/B}.\label{eq:
z star}
\end{equation}
We have
\begin{equation}
R(0)=\lim_{W_c\rightarrow \infty} R_B(1/W_c) = \sup_{0\le\rho\le
1} -\frac{z}{\rho}+\frac{1}{T_c}\frac{B\ln\left(1+\frac{\rho
PT_c}{B(1+\rho)}\right)}{\rho},\label{eq: D 00 B}
\end{equation}
and
\begin{equation}
\dot{R}(0)=-\frac{P^2}{B(1+\rho)(1+\rho^*+ \frac{\rho^*
PT_c}{B})^2},
\end{equation}
where $\rho^*$ is the optimizing $\rho$ in (\ref{eq: D 00 B}).
\hfill $\diamond$
\end{theorem}

The constraint on $z$ in (\ref{eq: fec}) again comes from the fact
that the reliability function is only known when $R\ge R_{crit}.$
Now we show that $z^*$ given by (\ref{eq: z star}) is the
corresponding error exponent at $R_{crit}$ when $W_c$ goes to
infinity. From the property of the critical rate $R_{crit},$ we
know the optimizing $\rho$ in (\ref{eq: D 00 B}) at the
corresponding error exponent $z_{crit}$ is $1.$ Thus, taking
derivative of the right side of (\ref{eq: D 00 B}) with respect to
$\rho,$ we must have
$$
\frac{z_{crit}}{\rho^2}-\frac{B}{T_c} \frac{\ln(1+\frac{\rho
PT_c}{B(1+\rho)})}{\rho^2} +\frac{B}{T_c}\frac{PT_c/B}{\rho(1+
\frac{\rho PT_c}{B(1+\rho)})} \frac{1}{(1+\rho)^2} |_{\rho=1}=0.
$$
By solving this, it is straightforward to have $z_{crit}=z^*$ with
$z^*$ determined by (\ref{eq: z star}). The corresponding rate
$R_{crit}$ can be obtained as follows
\begin{eqnarray*}
R_{crit}&=& -z_{crit}+\frac{B}{T_c}\ln(1+\frac{PT_c}{2B}) \\
&=&\frac{P}{4+2\frac{PT_c}{B}}.
\end{eqnarray*}
Using a similar argument as in the AWGN channel case, we can argue
that for $z\in(0,z^*),$ the reliability function coincides with
the random-coding exponent for sufficiently large $W_c$ . Thus,
the calculation of $R(0)$ and $\dot{R}(0)$ can be carried out by
using the random-coding exponent.

\begin{figure}
\center \epsfig{file=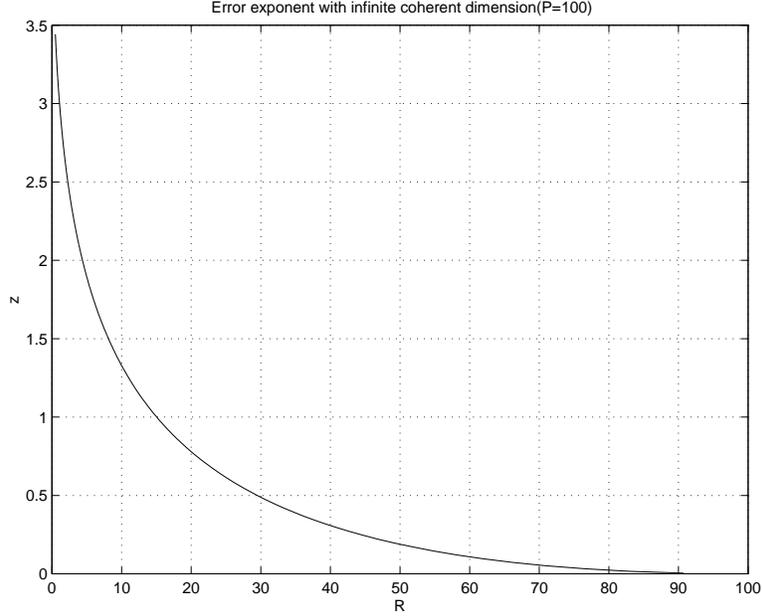,width=4.0in} \caption{The error
exponent curve from $R_{crit}$ to capacity for the channel with
infinite coherence dimension. $B=T_c=1.$ $P=100.$}\label{fig:
infinite}
\end{figure}

Another observation here is that the applicable region (in terms
of $R$), where the random-coding exponent coincides with the
sphere-packing exponent, actually covers most of the rate region
from $0$ to capacity, when the available energy per coherence
block $\frac{PT_c}{B}$ is fairly large. To see this, we first
notice that as $W_c$ goes to infinity, our capacity $C_{\infty}$
in (\ref{eq: fc}) is $P.$ Thus, the critical rate $R_{crit}$ can
be also written as $\frac{1}{4+2\frac{PT_c}{B}}C_{\infty}.$ When
$\frac{PT_c}{B}$ is large, we have $R_{crit}<<C_{\infty}.$ This
observation is also shown in Figure~\ref{fig: infinite}. For
simplicity, we choose $B=T_c=1$ in this numerical example and
choose $P=100.$

Next, we need to identify those signaling schemes which can
achieve $R(0)$ and $\dot{R}(0).$ Again, we consider BPSK and QPSK
signaling. However, for the fading channel (\ref{eq: fading vector
channel}), these two signaling schemes have slightly different
meanings than what we defined in last section for AWGN channels.
Specifically, for both BPSK and QPSK, we spread the available
power in each coherent block equally among all the time-frequency
coherent blocks and make the distributions in each dimension
i.i.d. For BPSK, the symbols for each dimension are
$\sqrt{P/BW_c}$ and $-\sqrt{P/BW_c},$ with equal probability. For
QPSK, the symbols are $\sqrt{\frac{P}{2BW_c}}(1+j)$,
$\sqrt{\frac{P}{2BW_c}}(1-j)$, $\sqrt{\frac{P}{2BW_c}}(-1+j)$ and
$\sqrt{\frac{P}{2BW_c}}(-1-j).$ Similar to the AWGN case, we have

\begin{theorem}\label{thm: BPSK and QPSK DB}
Both BPSK and QPSK are first-order optimal for any given
$z\in(0,z^*)$; however, only QPSK is second-order optimal. \hfill
$\diamond$
\end{theorem}

\subsection{Implications and discussion}

The results that we have obtained for both AWGN channels and
coherent fading channels are consistent with the results from a
capacity point of view in the seminal work \cite{ver02}. By
letting $z$ go to $0,$ the quantity $R_z$ becomes the capacity of
the channel. Thus, it can be easily checked that by taking $z$ to
be $0,$ we can recover the capacity results by using the
expressions in Theorem~\ref{thm: non-fading} and Theorem~\ref{thm:
fading D B}. However, we also have to point out that in
\cite{ver02}, a very general treatment is provided for a much
broader class of channel models. In this paper, due to the
complexity of the calculation of the reliability function, we only
calculated the first and second order rate approximation for two
very specific channel models.

Despite the similarity between our results and Verdu's results
regarding near-optimal signaling, the fact that QPSK is {\em
still} near-optimal under a certain error exponent constraint is
still somewhat surprising because of the following reason. In
general, very little is known about the conditions under which an
input distribution achieves the optimal error exponent at a given
rate, even in the infinite bandwidth limit. It is not necessarily
true that capacity-achieving distributions are also optimal from
an error-exponent point of view. One example is the
infinite-bandwidth non-coherent Rayleigh fading channel, which is
studied in \cite{wusri04}. Thus, it is not obvious that actually
QPSK can do well in the wideband regime from an error exponent
point of view,
even though it is wideband optimal from a capacity
point of view.

\section{Proof of Theorem~\ref{thm: non-fading} and
Theorem~\ref{thm: BPSK and QPSK}}\label{sec: AWGN proof}

Due to the technical nature of the calculations needed in the
proofs of our main results, we first summarize the proof steps as
follows to help the reader follow the proof of our main results.

The proof of Theorem~\ref{thm: non-fading} can be broken down into
the following major steps:
\begin{itemize}
\item[1.] We first relate the problem of finding $R(0)$ and
$\dot{R}(0),$ where $R$ is the communication rate per second as a
function of $1/B,$ to the problem of finding $\dot{r}(0)$ and
$\ddot{r}(0),$ where $r$ is the communication rate per degree of
freedom in (\ref{eq: scalar channel}) as a function of $p,$ which
denotes the SNR per degree of freedom.
\item[2.] The calculation of $\dot{r}(0)$ can be related to the
optimal value for $E_o$ in the infinite bandwidth limit; an upper
bound is derived for $E_o$ using a simple inequality; this bound
is further shown to be achievable;

\item[3.]$\ddot{r}(0)$ can also be related to certain derivatives
of $E_o$; a better upper bound is derived for $E_o$ which yields
an upper bound for $\ddot{r}(0);$ this bound is also shown to be
achievable.
\end{itemize}

The next several subsections will prove the main results following
these three steps.

\subsection{Communication rate and error exponent per degree of freedom}

It is shown in \cite{ver02} that the capacity $C$ in a bandlimited
channel with limited available power $P,$ but large available
bandwidth $B,$ can be related to the capacity $c$ in a scalar
channel with small available power $p=P/B.$ Thus, the problem of
finding optimal $C(0)$ and $\dot{C}(0)$ can be shown to be
equivalent to the problem of finding optimal $\dot{c}(0)$ and
$\ddot{c}(0)$. The relationship between $C(0)$ and $\dot{c}(0)$ is
also extensively studied in an earlier paper \cite{ver90}, where
the notion {\sl capacity per unit cost} was studied. We first show
that a similar connection can be made between the error-exponent
constrained rates $R$ (nats per second) and $r$ (nats per symbol).

\begin{theorem}\label{thm: R and r}
Consider a scalar Gaussian channel $y=x+w$ with average power
constraint $p.$ Further, the signaling schemes are constrained by
${\tilde \mathcal F}(p)=\left\{ \left\{q_p({x})\right\}:q_p({
x})\in {\mathcal Q}(p)\right\}.$ Let $r$ be the maximum rate per
symbol at which information can be transmitted through channel
(\ref{eq: scalar channel}) such that the error exponent satisfies
$$
\hat{E}(r,p)\ge pz,\quad 0<z<\frac{1}{4},
$$
where $\hat{E}(r,p)$ is the error exponent per symbol of the
scalar channel with power constraint $p.$ Consider $r$ as a
function of $p.$ Let $R$ (nats per second) be defined as the
solution to (\ref{eq: error constraint}). We have
\begin{eqnarray*}
R(0)&=&P\dot{r}(0);\\
\dot{R}(0)&=&\frac{P^2\ddot{r}(0)}{2}.
\end{eqnarray*}
\end{theorem}
Proof: It is easy to check that 
$$
E(R,P,B)=B\hat{E}(R/B,P/B).
$$
Denoting $r=R/B$ and $p=P/B,$ the original error-exponent
constraint can be rewritten as
$$
\hat{E}_r(r,p)\ge pz.
$$
Using these two relations and considering $R$ as a function of
$b=1/B,$ we have
\begin{eqnarray}
R(0)&=&\lim_{b\rightarrow 0} R(b)=\lim_{b\rightarrow 0}
\frac{r(p)}{b}=P\lim_{b\rightarrow
0}\frac{r(p)}{p}=P\dot{r}(0)\label{eq: first relation}\\
\dot{R}(0)&=&\lim_{b\rightarrow 0} \frac{R(b)-R(0)}{b} =\lim_{b
\rightarrow 0} \frac{\frac{1}{b}r(Pb)-R(0)}{b}
=\frac{P^2\ddot{r}(0)}{2} \label{eq: second relation}
\end{eqnarray} \hfill $\diamond$

Thus, the original problem of finding $R(0)$ and $\dot{R}(0)$ in
the wideband regime is equivalent to finding the optimal values
for $\dot{r}(0)$ and $\ddot{r}(0),$ given a constraint on the
reliability function $ \hat{E}(r,p)\ge pz.$ In the rest of this
paper, we will deal with this scalar channel problem. For
notational convenience, we use $E(r,p)$ to denote the error
exponent per symbol of the single channel instead of using ${\hat
E}(r,p).$

\subsection{Optimal value of $\dot{r}(0)$}

We know for the error-exponent constraint in the range of
$(0,\frac{1}{4})$ and $p$ sufficiently small, we have
$$
E(r,p)=E_r(r,p)=\sup_{0\le\rho\le 1} -\rho r+E_o(p,\rho),
$$
where
\begin{equation}
E_o(p,\rho)=\sup_{q_p\in{\tilde\mathcal F}(p)}\sup_{\beta\ge 0}
-\ln \int \left(\int q_p(x) e^{\beta(|x|^2-p)}
f(y|x)^{\frac{1}{1+\rho}} dx \right)^{1+\rho} dy.
\end{equation}
Thus, the constraint on the error exponent can also be written as
\begin{equation}
pz=\sup_{0\le\rho\le 1} -\rho r+E_o(p,\rho).\label{eq: constraint}
\end{equation}

The first result in the first-order calculation is the following
lemma.
\begin{lemma}\label{lem: upper bound}
For any $\rho\in[0,1],$ $E_o(p,\rho)$ is upper bounded by
\begin{equation}
E_o(p,\rho)\le \frac{p\rho}{1+\rho}.
\end{equation}
\end{lemma}
Proof: For notational convenience, define $\alpha(y)$ to be
\begin{equation}
\alpha(y)=\int q_p(x) e^{\beta(|x|^2-p)} f(y|x)^{\frac{1}{1+\rho}}
dx
\end{equation}
and $M(y)$ as
\begin{equation}
M(y)=\int q_p(x) e^{\beta(|x|^2-p)}
\left[\frac{f(y|x)}{f(y|0)}\right]^{\frac{1}{1+\rho}}
dx.\label{eq: def M 0}
\end{equation}
Here $f(y|0)$ denotes the distribution function of $y$ conditioned
on that the input is $0.$ It is easy to see that $f(y|0)$ is simply the
distribution of the Gaussian noise $f_w(y).$ Then we have
\begin{eqnarray}
E_o(p,\rho)&=&\sup_{q\in {\tilde\mathcal F}(p)}\sup_{\beta\ge 0}
-\ln \int \alpha(y)^{1+\rho} dy\nonumber\\
&=&\sup_{q\in {\tilde\mathcal F}(p)}\sup_{\beta\ge 0} -\ln\int
f_w(y) M(y)^{1+\rho} dy \\
&\le& \sup_{q\in {\tilde\mathcal F}(p)}\sup_{\beta\ge 0}
-\ln\left(\int f_w(y) M(y) dy\right)^{1+\rho}\label{eq: b0}\\
&=&\sup_{q\in {\tilde\mathcal F}(p)}\sup_{\beta\ge 0} -(1+\rho)\ln
E_q\left[e^{\beta(|x|^2-p)} \int f_w(y)^\frac{\rho}{1+\rho}
f(y|x)^\frac{1}{1+\rho} dy \right]\\
&=&\sup_{q\in {\tilde\mathcal F}(p)}\sup_{\beta\ge 0} -(1+\rho)\ln
E_q\left[e^{\beta(|x|^2-p)} \int f_w(y)^\frac{\rho}{1+\rho}
f_w(y-x)^\frac{1}{1+\rho} dy \right]\\
&=& \sup_{q\in {\tilde\mathcal F}(p)}\sup_{\beta\ge 0}
-(1+\rho)\ln E_q\left[e^{\beta(|x|^2-p)} e^{-\theta
|x|^2}\right] \label{eq: b02}\\
&\le& \sup_{q\in {\tilde\mathcal F}(p)}\sup_{\beta\ge 0}
-(1+\rho)\ln  e^{-\theta p} \label{eq: b1} \\
&=&\frac{\rho p}{1+\rho}\nonumber,
\end{eqnarray}
where $\theta$ in (\ref{eq: b02}) is defined by
$$
\theta=\frac{\rho}{(1+\rho)^2}.
$$
The inequalities in (\ref{eq: b0}) and (\ref{eq: b1}) are simple
applications of Jensen's inequality. \hfill $\diamond$

The next theorem establishes an alternate expression for the error
exponent constraint (\ref{eq: constraint}).
\begin{theorem}\label{thm: equivalent form}
The error-exponent constraint (\ref{eq: constraint}) implies the
following relationship between $r$ and $z$
\begin{equation}
r=\sup_{0\le\rho\le 1} -\frac{pz}{\rho}+\frac{E_o(p,\rho)}{\rho}.
\label{eq: equivalent form}
\end{equation}
\end{theorem}
Proof: See Appendix~\ref{sec: equivalent form}. \hfill $\diamond$

Since we want to study the first and second-order derivative of
$r$ with respect to $p$ in the low SNR regime, it is more
convenient to use (\ref{eq: equivalent form}). To obtain the first
order derivative, from (\ref{eq: equivalent form}) we first note
that
$$
\frac{r}{p}=\sup_{0\le\rho\le 1}
-\frac{z}{\rho}+\frac{E_o(p,\rho)}{p\rho}.
$$
Now we relate $\dot{r}(0)$ to the first partial derivative of
$E_o(p,\rho)$ with respect to $p.$
\begin{theorem}\label{thm: r dot 0}
If as $p\rightarrow 0,$ the limit of $\frac{E_o(p,\rho)}{p}$ exists
for any $\rho\in[0,1],$ which is denoted as $\dot{E}_o(0,\rho),$ and further,
$$
\frac{E_o(p,\rho)}{p\rho}\rightarrow
\frac{\dot{E_o}(0,\rho)}{\rho} \quad\textsl{uniformly for
$\rho\in[0,1],$}
$$
we have
\begin{eqnarray}
\dot{r}(0)=\sup_{0\le \rho\le 1}
-\frac{z}{\rho}+\frac{\dot{E}_o(0,\rho)}{\rho}.\label{eq: r dot 0}
\end{eqnarray}
\end{theorem}
Proof: From the definition of {\em uniform convergence}, for any
$\epsilon>0,$ we can find $\delta(\epsilon)>0,$ such that for any
$p<\delta(\epsilon),$ we have
$$
\left|\frac{E_o(p,\rho)}{p\rho}-\frac{\dot{E_o}(0,\rho)}{\rho}\right|<
\epsilon, \quad \forall \rho\in[0,1].
$$
Thus, if we denote $K=\sup_{0\le \rho\le 1}
-\frac{z}{\rho}+\frac{\dot{E}_o(0,\rho)}{\rho},$ we have
$$
\frac{r(p)}{p}\le \sup_{0\le\rho\le 1} -\frac{z}{\rho}+
\frac{\dot{E_o}(0,\rho)}{\rho}+\epsilon=K+\epsilon.
$$
Similarly, we can show that $\frac{r(p)}{p}\ge K-\epsilon.$
Letting $\epsilon\rightarrow 0,$ we have
$\dot{r}(0)=\lim_{p\rightarrow 0}\frac{r(p)}{p}=K.$ \hfill
$\diamond$

\begin{lemma}\label{lem: uni1}
As $p\rightarrow 0,$ $\frac{E_o(p,\rho)}{p\rho}$ converges to
$\frac{1}{1+\rho}$ uniformly for $\rho\in[0,1].$
\end{lemma}
Proof: In Lemma~\ref{lem: upper bound}, we have already shown that
$$
\frac{E_o(p,\rho)}{p\rho}\le \frac{1}{1+\rho}.
$$
In Appendix~\ref{sec: BPSK and QPSK}, we will show that when the
input distribution is chosen to be BPSK or QPSK, $\frac{{\tilde
E}_o(p,q_p,\rho)}{p\rho}$ converges uniformly to
$\frac{1}{1+\rho}.$ Since $\frac{E_o(p,\rho)}{p\rho}$ is lower
bounded by $\frac{{\tilde E}_o(p,q_p,\rho)}{p\rho},$ the lemma
follows. \hfill $\diamond$

Using Lemma~\ref{lem: uni1} and Theorem~\ref{thm: r dot 0}, we can
compute $\dot{r}(0).$
\begin{proposition} \label{prop: dot r 0}
For $0< z< \frac{1}{4},$
\begin{equation}
\dot{r}(0)= (1-\sqrt{z})^2. \label{eq: dot r bound 0}
\end{equation}
\end{proposition}
Proof: From Theorem~\ref{thm: r dot 0}, we have
\begin{eqnarray}
\dot{r}(0)&=& \sup_{0\le\rho\le 1} -\frac{z}{\rho} +
\frac{\dot{E}_o(0,\rho)}{\rho} \label{eq: temp1}
\\
&=& \sup_{0\le\rho\le 1} -\frac{z}{\rho} +
\frac{1}{1+\rho}\nonumber\\
&=&\left\{
\begin{array}{ll}
(1-\sqrt{z})^2&0\le z\le\frac{1}{4};\\
\frac{1}{2}-z&\frac{1}{4}\le z\le 1.\\
\end{array}
\right.
\end{eqnarray}
For $0<z<\frac{1}{4},$ the optimizing
$\rho^*=\frac{\sqrt{z}}{1-\sqrt{z}}.$ \hfill $\diamond$

Note here the optimal value $\dot{r}(0)$ is obtained by optimizing
over all input distributions in ${\tilde\mathcal F}(p).$ However,
this result is valid for all input distributions. In other words,
allowing continuous alphabet or peaky signaling would not change
this optimal value. This is due to the well-known infinite
bandwidth AWGN channel error-exponent result, which is shown in
(\ref{eq: infinite bandwidth E}). It can be easily seen that
(\ref{eq: dot r bound 0}) is simply the inverse function of
(\ref{eq: infinite bandwidth E}). The purpose of deriving
$\dot{r}(0)$ using the constraint ${\tilde\mathcal F}(p)$ is not
to just derive (\ref{eq: dot r bound 0}), but also to obtain
conditions on the input distributions in ${\tilde\mathcal F}(p)$
which achieve (\ref{eq: dot r bound 0}). We will obtain such
conditions in the next subsection.

\subsection{First-order optimality condition}

Next we study conditions for a sequence of input distributions to
be first-order optimal.

\begin{lemma}\label{lem: first order condition 1 0}
Assuming $0<z<\frac{1}{4}$, a sufficient condition for
$\{q_p\}\in{\tilde\mathcal F}(p)$ to be first-order optimal is
that
\begin{equation}
\lim_{p\rightarrow 0} \frac{{\tilde
E}_o(p,q_p,\rho^*)}{p}=\frac{\rho^*}{1+\rho^*},\label{eq: first
order condition}
\end{equation}
where $\rho^*=\frac{\sqrt{z}}{1-\sqrt{z}}$.
\end{lemma}
Proof: If $\lim_{p\rightarrow 0} \frac{{\tilde
E}_o(p,q_p,\rho^*)}{p}=\frac{\rho^*}{1+\rho^*},$ we have
\begin{eqnarray*}
\liminf_{p\rightarrow 0} \frac{\tilde{r}}{p}&\ge&
\liminf_{p\rightarrow 0}
-\frac{z}{\rho^*}+ \frac{{\tilde E}_o(p,q_p,\rho^*)}{p\rho^*}\\
&=& -\frac{z}{\rho^*}+\lim_{p\rightarrow 0} \frac{{\tilde
    E}_o(p,q_p,\rho^*)}{p\rho^*}\\
&=& -\frac{z}{\rho^*}+\frac{1}{1+\rho^*}\\
&=& (1-\sqrt{z})^2.
\end{eqnarray*}
On the other hand, from Lemma~\ref{lem: upper bound}, we know
\begin{eqnarray*}
\limsup_{p\rightarrow 0} \frac{\tilde{r}}{p}
&=&\limsup_{p\rightarrow 0} \sup_{0\le \rho\le 1} -\frac{z}{\rho}
+\frac{{\tilde E}_o(p,q_p,\rho)}{p\rho} \\
&\le& \limsup_{p\rightarrow 0} \sup_{0\le \rho\le 1}
-\frac{z}{\rho} +\frac{E_o(p,\rho)}{p\rho}\\
&\le& \limsup_{p\rightarrow 0} \sup_{0\le \rho\le 1}
-\frac{z}{\rho} +\frac{1}{1+\rho}\\
&=& (1-\sqrt{z})^2.
\end{eqnarray*}
Thus, the limit of $\frac{\tilde r}{p}$ exists and we have
$$
\dot{\tilde{r}}(0)=\lim_{p\rightarrow 0} \frac{\tilde{r}}{p} =
(1-\sqrt{z})^2.
$$
\hfill $\diamond$

Actually, it does not take much to be first-order optimal.
\begin{lemma}\label{lem: mean zero optimal}
For a fixed $0<z<\frac{1}{4},$ a sequence of input distribution
${q_p}\in{\tilde\mathcal F}(p)$ is first-order optimal if it is
symmetric around $0.$
\end{lemma}
Proof: Refer to Appendix~\ref{sec: proof all}. \hfill $\diamond$

\subsection{The optimal value of $\ddot{r}(0)$}

In this section, we will find an upper bound for $\ddot{r}(0)$ and
later we will show that this value is achievable. To do this, we
first connect $\ddot{r}(0)$ to the second partial derivative of
$E_o(p,\rho)$ with respect to $p.$

\begin{theorem}\label{thm: r ddot 0}
Assume the second partial derivative of $E_o(p,\rho)$ with respect
to $p$ at $p=0$ (denoted as $\ddot{E}_o(0,\rho)$) exists for any
$\rho\in[0,1].$ Further, assume that
$$
\frac{\frac{E_o(p,\rho)}{p\rho}-\frac{\dot{E}_o(0,\rho)}{\rho}}{p}\rightarrow
\frac{\ddot{E}_o(0,\rho)}{2\rho} \quad \textsl{uniformly for $\rho\in[0,1]$},
$$
and $\frac{\ddot{E}_o(0,\rho)}{\rho}$ is a continuous and bounded
function of $\rho$ for $\rho\in[0,1].$ Then $\ddot{r}(0)$ can be
determined by
\begin{equation}
\ddot{r}(0)=\frac{\ddot{E}_o(0,\rho^*)}{\rho^*}, \label{eq: r ddot
0}
\end{equation}
where $\rho^*$ is the optimal $\rho$ in (\ref{eq: r dot 0}) and is
equal to $\frac{\sqrt{z}}{1-\sqrt{z}}.$
\end{theorem}
Proof: First we show that
$$\ddot{\overline{r}}(0)=\limsup_{p\rightarrow 0}
\frac{r(p)-p\dot{r}(0)} {p^2/2} \le
\frac{\ddot{{E}}_o(0,\rho^*)}{\rho^*}.$$

The uniform convergence gives us: for any $\epsilon>0,$ we can
find $\eta(\epsilon)$ such that for all $p<\eta(\epsilon),$
$$
\left|\frac{\frac{E_o(p,\rho)}{p\rho}-\frac{\dot{E}_o(0,\rho)}{\rho}}{p}-
\frac{\ddot{E}_o(0,\rho)}{2\rho}\right|<\epsilon \quad\textsl{for all
  $\rho\in[0,1]$}.
$$
In other words, for $p<\eta(\epsilon),$ we can write
$$
E_o(p,\rho)\le \dot{{E}}_o(0,\rho)p+\ddot{E}_o(0,\rho)p^2/2+
\rho\epsilon p^2.
$$
From (\ref{eq: equivalent
form}), we have
\begin{equation}\label{eq: temp020}
r(p) \le \sup_{0\le \rho\le 1}
-\frac{pz}{\rho}+\frac{\dot{{E}}_o(0,\rho)p
+\ddot{E}_o(0,\rho)p^2/2}{\rho}+ \epsilon p^2.
\end{equation}

Assume $\rho(p)$ is the optimizing $\rho$ for (\ref{eq: temp020}).
From the first-order calculation, we already know that
$$
\dot{E}_o(0,\rho)=\frac{\rho}{1+\rho}.
$$
Since the optimization in (\ref{eq: temp020}) is performed over a
compact set $[0,1]$ and by assumption $\ddot{E}_o(0,\rho)$ is
continuous in $\rho,$ the optimizing $\rho$ must exist.

We must have
\begin{eqnarray*}
r(p)&\le&\left\{\sup_{0\le\rho\le
1}-\frac{pz}{\rho}+\frac{p\dot{{E}}_o(0,\rho)}{\rho}\right\}+
\frac{\ddot{{E}}_o(0,\rho(p))\frac{p^2}{2}}{\rho(p)} +\epsilon p^2.
\end{eqnarray*}
From (\ref{eq: r dot 0}), we know
$$
\dot{r}(0)p= \sup_{0\le\rho\le
1}-\frac{pz}{\rho}+\frac{p\dot{{E}}_o(0,\rho)}{\rho}.
$$
This gives us
$$
\frac{r(p)-p\dot{r}(0)}{p^2/2} \le
\frac{\ddot{{E}}_o(0,\rho(p))}{\rho(p)}+ 2\epsilon.
$$
Letting $\epsilon$ go to $0$, we have
\begin{eqnarray}
\ddot{\overline{r}}(0)&=&\limsup_{p\rightarrow 0}
\frac{r(p)-p\dot{r}(0)}{p^2/2} \nonumber \\
&\le& \limsup_{p\rightarrow 0}
\frac{\ddot{{E}}_o(0,\rho(p))}{\rho(p)}\nonumber\\
&=& \frac{\ddot{{E}}_o(0,\rho^*)}{\rho^*},\label{eq: p00}
\end{eqnarray}
where $\rho^*$ is the optimizing $\rho$ of (\ref{eq: temp020}) as
$p$ goes to zero, and can be shown to be equal to
$\frac{\sqrt{z}}{1-\sqrt{z}}.$ The last equation (\ref{eq: p00})
can be easily verified given that
$\frac{\ddot{E}_o(0,\rho)}{\rho}$ is a continuous function of
$\rho,$ if we have $\lim_{p\rightarrow 0} \rho(p)=\rho^*,$ which
we will show in Appendix~\ref{sec: limit}.

To complete the proof of the theorem, it suffices to show
$$\ddot{\underline{r}}(0)= \liminf_{p\rightarrow 0}
\frac{r(p)-p\dot{r}(0)} {p^2/2}\ge
\frac{\ddot{{E}}_o(0,\rho^*)}{\rho^*}.$$

To see this, we choose $\rho=\rho^*$ in (\ref{eq: temp020}) and we
have
\begin{eqnarray*}
r(p)
&\ge&-\frac{pz}{\rho^*}+\frac{p\dot{{E}}_o(0,\rho^*)}{\rho^*}+
\frac{\ddot{{E}}_o(0,\rho^*)\frac{p^2}{2}}{\rho^*} -\epsilon p^2.
\end{eqnarray*}
From (\ref{eq: r dot 0}), we must have
$$
\dot{{r}}(0)=-\frac{z}{\rho^*}+\frac{\dot{{E}}_o(0,\rho^*)}{\rho^*},
$$
and thus, we have
\begin{eqnarray*}
\frac{r(p)-p\dot{r}(0)} {p^2/2}&\ge&
\frac{\ddot{E}_o(0,\rho^*)}{\rho^*} -2\frac{\epsilon p^2}{p^2}.
\end{eqnarray*}

Letting $p\rightarrow 0$, we will have
$$
\ddot{\underline{r}}(0)\ge \frac{\ddot{{E}}_o(0,\rho^*)}{\rho^*}.
$$
\hfill $\diamond$

Thus, to obtain the optimal value for $\ddot{r}(0)$, we need to
verify the uniform convergence assumption in Theorem~\ref{thm: r
ddot 0} and calculate $\frac{\ddot{E}_o(0,\rho^*)}{\rho^*}.$ To
show uniform convergence, we both upper and lower bound
$$\frac{\frac{E_o(p,\rho)}{p\rho}-\frac{\dot{E}_o(0,\rho)}{\rho}}{p}$$
by a function of $\rho$ plus a small term $\delta(1),$ which
converges to $0$ uniformly for $\rho\in[0,1],$ as $p$ goes to $0.$
Specifically, we want to show that when $p$ is small, we have
$$
\frac{\ddot{E}_o(0,\rho)}{2\rho}+\delta_1(1)\le
\frac{\frac{E_o(p,\rho)}{p\rho}-\frac{\dot{E}_o(0,\rho)}{\rho}}{p}
\le \frac{\ddot{E}_o(0,\rho)}{2\rho}+\delta_2(1),
$$
where both $\delta_1(1)$ and $\delta_2(1)$ converge to $0$
uniformly as $p$ goes to $0.$ The uniform convergence of
$$\frac{\frac{E_o(p,\rho)}{p\rho}-\frac{\dot{E}_o(0,\rho)}{\rho}}{p}$$
follows easily from here. We will first show an upper bound, then
we will obtain a lower bound by using
QPSK signaling at the input. In the rest of the paper, we will use the notation
$\delta(p^m)$ to denote a term satisfying that as $p$ goes to $0,$
$\frac{\delta(p^m)}{p^m}\rightarrow 0$ uniformly for
$\rho\in[0,1].$

We know that
$$
E_o(p,\rho)=\sup_{\{q_p\}\in {\tilde\mathcal F}(p)} {\tilde
E}_o(p,q_p,\rho).
$$
However, it is easy to see that we will not lose any optimality if
we constraint ourselves to those input distributions which perform
at least as good as QPSK. In other words, we have
\begin{equation}
E_o(p,\rho)=\sup_{\{q_p\}\in \tilde{\mathcal G}(p)} {\tilde
E}_o(p,q_p,\rho),
\end{equation}
where $\tilde{\mathcal G}(p)$ is defined as
\begin{equation}
\tilde{\mathcal G}(p)=\left\{ \{q_p\}\in {\tilde\mathcal F}(p):
{\tilde E}_o(p,q_p,\rho)\ge {\tilde E}_o(p,QPSK,\rho), \forall p>0
\right\}
\end{equation}

\begin{lemma}\label{lem: temp18 0}
For any sequence of input distributions $\{q_p(x)\}\in {\tilde \mathcal G}(p)$,
\begin{equation}\label{eq: temp18 0}
\frac{\frac{E_o(p,\rho)}{p\rho}-\frac{\dot{E}_o(0,\rho)}{\rho}}{p}\le
\frac{-\inf_{\{q_p\}\in \tilde{\mathcal G}(p)}\inf_{\beta \ge
0}\int \alpha(y)^{1+\rho}dy +e^{-\frac{\rho p}{1+\rho}} }{\rho
p^2}.
\end{equation}
\end{lemma}
Proof: See Appendix~\ref{sec: temp18 0}. \hfill $\diamond$

Next, we further bound $\int \alpha(y)^{1+\rho} dy$ for any sequence
of input distributions $\{q_p\}\in \tilde{\mathcal G}(p).$
\begin{lemma}
For all $q_p(x)$ and all $\beta$, we have
\begin{eqnarray}
\int \alpha(y)^{1+\rho}dy &=& \int f_w(y) (1+T(y))^{1+\rho} dy\nonumber\\
&\ge& 1+(1+\rho)\int f_w(y)T(y)dy+\frac{\rho(1+\rho)}{2}\int
f_w(y)T^2(y) dy+\frac{\rho(1+\rho)(\rho-1)}{6}\int f_w(y)T^3(y)
dy,\nonumber\\
\label{eq: second order inequ 0}
\end{eqnarray}
where $T(y)=M(y)-1$ and $M(y)$ is defined by (\ref{eq: def M 0}).
\end{lemma}
Proof: The following inequality is true for all $t\ge -1$ and all
$\rho \in [0,1]:$
$$
(1+t)^{1+\rho}\ge
1+(1+\rho)t+\frac{\rho(1+\rho)}{2}t^2+\frac{\rho(1+\rho)(\rho-1)}{6}t^3.
$$
Using the fact that $$\int\alpha(y)^{1+\rho}dy= \int
f_w(y)(1+T(y))^{1+\rho}dy$$ and plugging in the above inequality, we
have (\ref{eq: second order inequ 0}). \hfill $\diamond$

We will now treat the three terms separately in (\ref{eq: second order inequ
0}) and find a bound for each of them.

\begin{lemma}
\begin{equation}
\int f_w(y) T(y) dy\ge e^{-\theta p} -1,\label{eq: first term}
\end{equation}
where $\theta=\frac{\rho}{(1+\rho)^2}.$
\end{lemma}
Proof: It is easy to check
$$
\int f_w(y) T(y) dy=E[e^{\beta(|x|^2-p)} e^{-\theta |x|^2}]-1.
$$
Applying Jensen's inequality here, we get (\ref{eq: first term}).
\hfill $\diamond$

\begin{lemma}\label{lem: T2 0}
For any input distribution $\{q_p(x)\}\in {\tilde \mathcal G}(p),$
let $\beta^*$ be the optimizing $\beta,$ which maximizes
\begin{equation}\label{eq: int alpha}
\sup_{\beta\ge 0} -\ln \int \alpha(y)^{1+\rho} dy.
\end{equation}
We have
$$
\left. \int f_w(y)T^2(y) dy\right|_{\beta=\beta^*} \ge \theta^2
p^2+\frac{p^2}{(1+\rho)^4}+\delta(p^2).
$$
\end{lemma}
Proof: See Appendix~\ref{sec: T2 0}. \hfill $\diamond$

For those input distributions in $\tilde{\mathcal G}(p),$ the term
with integral over $T^3(y)$ actually does not contribute anything
to the second-order calculation, which is shown in the following
lemma.

\begin{lemma}\label{lem: T3 0}
Suppose that $\{q_p(x)\}\in {\tilde\mathcal G}(p).$ We have
$$
\left.\int f_w(y)T^3(y)dy\right|_{\beta=\beta^*} =\delta(p^2).
$$
\end{lemma}
Proof: See Appendix~\ref{sec: appendix 3}. \hfill $\diamond$

With these results, it is straightforward to show the required
uniform convergence.
\begin{proposition}\label{prop: ddot Enote}
\begin{equation}
\frac{\frac{E_o(p,\rho)}{p\rho}-\frac{\dot{E}_o(0,\rho)}{\rho}}{p}
\rightarrow -\frac{1}{2(1+\rho)^3}\quad \textsl{uniformly for
$\rho\in[0,1],$}\label{eq: m00}
\end{equation}
 as $p$ goes to $0.$
\end{proposition}
Proof: Combining Lemma~\ref{lem: T2 0} and Lemma~\ref{lem: T3 0},
we have
\begin{eqnarray*}
\int \alpha(y)^{1+\rho} dy &\ge&
1-\frac{\rho}{1+\rho}p+(1+\rho)\theta^2p^2/2+\frac{\rho(1+\rho)}{2}
\left(\theta^2 p^2+\frac{p^2}{(1+\rho)^4}\right)+\rho \delta(p^2)\\
&=& 1-\frac{\rho}{1+\rho}p+\frac{\rho^2
p^2}{2(1+\rho)^2}+\frac{\rho p^2}{2(1+\rho)^3}+\rho \delta(p^2).
\end{eqnarray*}

Applying Lemma~\ref{lem: temp18 0} here, we can obtain that
$$
\frac{\frac{E_o(p,\rho)}{p\rho}-\frac{\dot{E}_o(0,\rho)}{\rho}}{p}
\le -\frac{1}{2(1+\rho)^3}+\delta(p^2).
$$

Later, we will show that by choosing the input distribution to be
QPSK, we can establish a lower bound which has the same expression
as the upper bound. Thus, we know (\ref{eq: m00}) is true. \hfill
$\diamond$

Since we know $\rho^*=\frac{\sqrt{z}}{1-\sqrt{z}}$, the following
corollary is a direct consequence of Theorem~\ref{thm: r ddot 0}.
\begin{corollary} For $0<z<\frac{1}{4},$ we have
\begin{equation}
\ddot{r}(0)=-(1-\sqrt{z})^3.
\end{equation}
\end{corollary}\hfill $\diamond$

\subsection{BPSK and QPSK}

Combining the results regarding $\dot{r}(0)$ and $\ddot{r}(0)$ in
the previous subsections and Theorem~\ref{thm: R and r}, we have
proved Theorem~\ref{thm: non-fading}. Regarding Theorem~\ref{thm:
BPSK and QPSK}, the first part of the theorem is a direct
consequence of Lemma~\ref{lem: mean zero optimal}, which has
already been proved. For the second part of the Theorem regarding
BPSK and QPSK signaling, we can again do the calculations in a
scalar channel with small power as we have proceeded with the
proof of Theorem~\ref{thm: non-fading}. The calculations are
rather straightforward and we put the detailed proof of this part
in Appendix~\ref{sec: BPSK and QPSK} for completeness.

\section{Proof of Theorem~\ref{thm: fading D B} and Theorem~\ref{thm: BPSK and QPSK DB}}\label{sec: fading
proof}

In this section, we will prove Theorem~\ref{thm: fading D B} and
Theorem~\ref{thm: BPSK and QPSK DB}. For simplicity, we only prove
the case for $B=1,$ i.e., we focus on one of the $B$ parallel
channels in the channel model (\ref{eq: fading vector channel}).
The extension to the general case with $B$ parallel channels is
quite straightforward. Since $B=1,$ we drop the subscript of $l$
in (\ref{eq: fading vector channel}) and we have
\begin{equation}\label{eq: vector channel 1}
{\bf y}=H{\bf x}+{\bf w}.
\end{equation}
We assume the average power available in each block is $PT_c,$
i.e.,
\begin{equation}
E[\|{\bf x}\|^2]=PT_c.\label{eq: power constraint D}
\end{equation}
Thus, the energy per degree of freedom is $\frac{P}{W_c},$ which
is small when $W_c$ is large.

In this proof, we will use the results for AWGN channels
extensively. To avoid confusion in the notation, we will use a
superscript ``NF" (Non-Fading) to denote any quantity that was
computed for the AWGN channel.

\subsection{$R(0)$ and first-order optimal condition}

In the near capacity region ($R>R_{crit}$), where the
random-coding exponent and sphere-packing exponent are tight, the
reliability function constraint can be written as
$$
\sup_{0\le \rho\le 1} -\rho R +{E}_o(P,\rho, W_c) = z,
$$
and
\begin{equation}
{E}_o(P,\rho,W_c)= \frac{1}{T_c}\sup_{q\in{\mathcal F}_{W_c}(P)}
\sup_{\beta\ge 0} -\ln E_{H}\left[\int(\int q({\bf x})
e^{\beta(\|{\bf x}\|^2-PT_c)}f({\bf y}|{\bf x},
H)^{\frac{1}{1+\rho}} d{\bf x})^{1+\rho} d{\bf
y}\right].\label{eq: tilde Enote}
\end{equation}

Similar to the AWGN case, we first show that $E_o(P,\rho,W_c)$ is
always a bounded quantity.
\begin{lemma}\label{lem: upper bound D}
For any $\rho\in [0,1],$
\begin{equation}
0\le E_o(P,\rho,W_c)\le \frac{1}{T_c}\ln(1+\frac{\rho
PT_c}{1+\rho}).\label{eq: upper bound D}
\end{equation}
\end{lemma}
Proof: The lower bound is easy to show from (\ref{eq: tilde
Enote}), using a similar approach as in the AWGN case:
\begin{eqnarray}
T_c{E}_o(P,\rho,W_c)&\ge& \sup_{q\in{\mathcal F}_{W_c}(P)} -\ln
E_{H}\left[\int(\int q({\bf x}) f({\bf y}|{\bf x},
H)^{\frac{1}{1+\rho}} d{\bf x})^{1+\rho} d{\bf y}\right]\label{eq: t00}\\
&\ge& \sup_{q\in{\mathcal F}_{W_c}(P)} -\ln E_{H}\left[\int(\int
q({\bf x}) f({\bf y}|{\bf x}, H) d{\bf x}) d{\bf y}\right]\label{eq: t01}\\
&=& 0.\nonumber
\end{eqnarray}
The inequality in (\ref{eq: t00}) comes from taking $\beta=0$ and
the inequality in (\ref{eq: t01}) follows from Jensen's equality,
by noticing that $t^{1+\rho}$ is a convex function.

To show the upper bound, we move the two supremums inside the
expectation over $H:$
$$
T_c{E}_o(P,\rho,W_c)\le -\ln E_{H} \left[\inf_{q_{H}\in{\mathcal
F}_{W_c}(P)} \inf_{\beta_H\ge 0} \int(\int q_{H}({\bf x})
e^{\beta_H(\|{\bf x}\|^2-PT_c)}f({\bf y}|{\bf x},
H)^{\frac{1}{1+\rho}} d{\bf x})^{1+\rho} d{\bf y}\right].
$$
Now for each realization of $H,$ we choose the best $q_H({\bf x})
$ and $\beta$ to optimize the integrand in the equation above.
This is the same as finding the optimal $q({\bf x})$ and $\beta$
in an AWGN vector channel with a fixed gain $H.$ Thus, we do not
lose any optimality by choosing $q({\bf x})$ to be i.i.d. in all
components of the vector. Denote $q_H({\bf x})=\Pi_{l=1}^D
\hat{q}_H(x_l),$ and we have
\begin{eqnarray}
&&T_c{E}_o(P,\rho,W_c)\nonumber\\
&\le& -\ln E_{H} \left[\inf_{\hat{q}_{ H}(x)\in{\mathcal
F}(\frac{P}{W_c})} \inf_{\beta_H\ge 0} \left(\int(\int \hat{q}_{
H}(x) e^{\beta_H(| x|^2-\frac{P}{W_c})}f(y|x,H)^{\frac{1}{1+\rho}}
d{x})^{1+\rho}
d{y}\right)^D\right]\nonumber \\
&=& -\ln E_{H}  \left[e^{ \inf_{\hat{q}_{ H}(x)\in{\mathcal
F}(\frac{P}{W_c})} \inf_{\beta_H\ge 0}D\ln \left(\int(\int
\hat{q}_{H}(x) e^{\beta_H(|
x|^2-\frac{P}{W_c})}f(y|x,H)^{\frac{1}{1+\rho}} d{x})^{1+\rho}
d{y}\right)}\right]\nonumber \\
&=& -\ln E_{H}  \left[e^{ \inf_{\hat{q}_{H}(x)\in{\mathcal
F}(\frac{P}{W_c})} \inf_{\beta_H\ge 0}D\ln \left(\int(\int
\hat{q}_{H}(x) e^{\frac{\beta_H}{|H|^2}(|
H|^2|x|^2-\frac{P|H|^2}{W_c})}f_w(y-Hx)^{\frac{1}{1+\rho}}
d{x})^{1+\rho}
d{y}\right)}\right]\nonumber \\
&=& -\ln E_{H} \left[ e^{-D E_o^{NF}(\frac{P|H|^2}{W_c}, \rho)
}\right],\label{eq: a00}
\end{eqnarray}
where $E_o^{NF}(p,\rho)$ denotes the $E_o$ for a scalar non-fading
(AWGN) channel,
$$
E_o^{NF}(p,\rho)=\sup_{\hat{q}({\bf x})\in {\mathcal F}(p)}
\sup_{\beta \ge 0} -\ln \int(\int \hat{q}(x) e^{\beta(\|
x\|^2-\frac{P}{D})}f_w(y-x)^{\frac{1}{1+\rho}} d{x})^{1+\rho}
d{y}.
$$
Here $f_w$ denotes the probability density function of a symmetric
complex Gaussian random variable with unit variance.

In last chapter, we have already shown that
$$
E_o^{NF}(p,\rho)\le \frac{p\rho}{1+\rho}.
$$
Plugging this into (\ref{eq: a00}), we get (\ref{eq: upper bound
D}).\hfill $\diamond$

With this upper bound, we can find the following equivalent form
of the error-exponent constraint, which is easier for us to work
with.
\begin{theorem}
An alternative form of the error-exponent constraint is
\begin{equation}
R(1/W_c)=\sup_{0\le \rho\le 1} -\frac{z}{\rho} +
\frac{{E}_o(P,\rho, W_c)}{\rho}.\label{eq: equivalent form D}
\end{equation}
\end{theorem}
Proof: Similar to the proof of Theorem~\ref{thm: equivalent
form}.\hfill $\diamond$

\begin{corollary}\label{cor: equivalent form}
In the equivalent form of the error-exponent constraint (\ref{eq:
equivalent form D}), we can restrict $\rho$ to be in interval
$[\frac{z}{P},1],$ without losing any optimality. In other words,
\begin{equation}
R(1/W_c)=\sup_{\frac{z}{P}\le \rho\le 1} -\frac{z}{\rho} +
\frac{{E}_o(P,\rho, W_c)}{\rho}.\label{eq: equivalent form 1}
\end{equation}
\end{corollary}
Proof: Note $R(1/W_c)$ is the maximum rate such that the
error-exponent constraint is satisfied. For a reasonable choice of
$z,$ (we will discuss later about the range of $z$ that we are
interested in,) the supremum in (\ref{eq: equivalent form D}) must
yield a non-negative result. Thus, we can restrict ourselves to
the $\rho$ such that $E_o(P,\rho,W_c)\ge z.$ Applying
Lemma~\ref{lem: upper bound D} here, this further implies
$$
\frac{1}{T_c}\ln(1+\frac{\rho PT_c}{1+\rho})\ge z.
$$
Noticing that $\ln(1+\frac{\rho PT_c}{1+\rho})\le \frac{\rho
PT_c}{1+\rho} \le \rho PT_c,$ we have $\rho P \ge z.$ Thus, we
only need to perform the optimization of $\rho$ in the interval
$[\frac{z}{P},1].$ \hfill $\diamond$

Since we are studying the behavior of $R(1/W_c)$ at large $W_c$
for a fixed $z>0,$ the range of $\rho$ in (\ref{eq: equivalent
form 1}) excludes $0,$ which will be quite helpful in the
calculations of $R(0)$ and $\dot{R}(0),$ as we will show later.

To find the value of $R(0)=\lim_{W_c\rightarrow \infty} R(1/W_c),$
an operation of exchanging the order of supremum and limit is
involved. We need the following theorem to justify this operation.
\begin{theorem}
If as $W_c$ goes to infinity, for any $\rho\in [0,1],$ the limit
of ${E}_o(P,\rho, W_c)$ exists, which is denoted as
$E_o(P,\rho,\infty),$ and further, ${E}_o(P,\rho, W_c)$ converges
to $E_o(P,\rho,\infty)$ uniformly for $\rho\in [0,1],$ we have
\begin{equation}
R(0)=\sup_{0\le \rho\le 1} -\frac{z}{\rho} + \frac{{E}_o(P,\rho,
\infty)}{\rho}.\label{eq: R 0}
\end{equation}
\end{theorem}
Proof: Uniform convergence of $E_o(P,\rho,W_c)$ gives us the
following: for any $\epsilon>0,$ we can find ${W_c}^{(\epsilon)},$
such that for any $W_c\ge W_c^{(\epsilon)},$ we have
$$
|E_o(P,\rho,W_c)-E_o(P,\rho,\infty)|\le \epsilon,\quad \textsl{for
all $\rho\in[0,1].$}
$$
From (\ref{eq: equivalent form 1}), we know for
$W_c>W_c^{(\epsilon)},$
\begin{eqnarray*}
R(1/W_c)&\le&\sup_{\frac{z}{P}\le \rho\le 1} -\frac{z}{\rho} +
\frac{{E}_o(P,\rho, \infty)+\epsilon}{\rho}\\
&\le& \sup_{\frac{z}{P}\le \rho\le 1} -\frac{z}{\rho} +
\frac{{E}_o(P,\rho, \infty)}{\rho}+\frac{P\epsilon}{z}.
\end{eqnarray*}
Similarly, we can show that
$$
R(1/W_c)\ge \sup_{\frac{z}{P}\le \rho\le 1} -\frac{z}{\rho} +
\frac{{E}_o(P,\rho, \infty)}{\rho}-\frac{P\epsilon}{z}.
$$
From here, it is easy to see that
$$
R(0)=\lim_{W_c\rightarrow \infty} R(1/W_c)= \sup_{\frac{z}{P}\le
\rho\le 1} -\frac{z}{\rho} + \frac{{E}_o(P,\rho, \infty)}{\rho}.
$$
The supremum over $[\frac{z}{P},1]$ and $[0,1]$ can be shown to be
equivalent using a similar argument as in the proof of
Corollary~\ref{cor: equivalent form}. Thus, (\ref{eq: R 0}) must
be true.\hfill $\diamond$

The uniform convergence can be easily established if we can find a
lower bound for $E_o(P,\rho,W_c)$ which converges to
$E_o(P,\rho,\infty)$ uniformly, since we have already obtained an
upper bound in Lemma~\ref{lem: upper bound D}. We will use a
widely-used signaling scheme, QPSK signaling, to establish a lower
bound for $E_o(P,\rho,W_c)$. Later, we will discuss the optimality
of QPSK and the lack of optimality of another widely used
signaling scheme, BPSK, in the wideband regime.

\begin{lemma}\label{lem: first Enote}
When the coherence dimension $W_c$ goes to infinity,
\begin{equation}
{E}_o(P,\rho,W_c)\rightarrow \frac{1}{T_c}\ln(1+\frac{\rho
PT_c}{1+\rho}) \enskip \textsl{uniformly for $\rho\in[0,1].$}
\end{equation}
\end{lemma}
Proof: Because of (\ref{eq: upper bound D}), it suffices to show
that for any $\epsilon>0,$ we can find $W_c^{(\epsilon)},$ such
that
$$
E_o(P,\rho,W_c)\ge \frac{1}{T_c}\ln(1+\frac{\rho PT_c}{1+\rho})
-\epsilon,
$$
for any $W_c\ge W_c^{(\epsilon)}$ and for all $\rho\in[0,1].$

From the definition of $E_o(P,\rho,W_c),$ we know for any specific
choice of $\{q^*\}\in {\mathcal F}_{W_c}(P),$ we have
$$
E_o(P,\rho,W_c)\ge {\tilde E}_o(P,q^*,\rho,W_c),
$$
where ${\tilde E}_o(P,q^*,\rho,W_c)$ is defined as follows
\begin{equation}
{\tilde E}_o(P,q^*,\rho,W_c)= \frac{1}{T_c}\sup_{\beta\ge 0} -\ln
E_{H}\left[\int(\int q^*({\bf x}) e^{\beta(\|{\bf
x}\|^2-PT_c)}f({\bf y}|{\bf x}, H)^{\frac{1}{1+\rho}} d{\bf
x})^{1+\rho} d{\bf y}\right]. \label{eq: def Enote qD}
\end{equation}

Now we choose $q^*$ to be QPSK. Since now $\|{\bf x}\|^2=PT_c$
with probability $1,$ the power-constraint parameter $\beta$ does
not affect ${\tilde E}_o(P,QPSK,\rho,W_c)$ and we have
\begin{eqnarray}
{\tilde E}_o(P,QPSK,\rho,W_c)=-\frac{1}{T_c}\ln E_H\left[ \exp\{-D
{\tilde
E}_o^{NF}(\frac{P|H|^2}{W_c},QPSK,\rho)\}\right],\label{eq: pr4}
\end{eqnarray}
where ${\tilde E}_o^{NF}(p,QPSK,\rho)$ is
$$
{\tilde E}_o^{NF}(p,QPSK,\rho)=-\ln \int
E_x[f_w(y-x)^\frac{1}{1+\rho}]^{1+\rho} dy.
$$

Next we show that for any $\epsilon>0,$ we can find
$W_c^{(\epsilon)},$ such that
$$
{\tilde E}_o(P,QPSK,\rho,W_c)\ge \frac{1}{T_c}\ln(1+\frac{\rho
PT_c}{1+\rho}) -\epsilon.
$$
From (\ref{eq: pr4}), it suffices to show that
\begin{equation}
(1+\frac{\rho P}{1+\rho}) E_H\left[ \exp\{-D {\tilde E}_o^{NF}
(\frac{P|H|^2}{W_c},QPSK,\rho)\}\right] <e^{\epsilon
T_c}.\label{eq: pr5}
\end{equation}

In last section, we have already shown that as $p\rightarrow 0,$
$\frac{{\tilde E}_o^{NF}(p,QPSK,\rho)}{p\rho}\rightarrow
\frac{1}{1+\rho}$ uniformly. In other words, for any
$\epsilon'>0,$ we can find $\xi>0,$ such that for all $p\le\xi,$
$$
\frac{{\tilde E}_o^{NF}
(p,QPSK,\rho)}{p\rho}>\frac{1}{1+\rho}-\epsilon',\quad \textsl{for
all $\rho\in[0,1],$}
$$
or equivalently,
\begin{equation}
{\tilde E}_o^{NF}(p,QPSK,\rho)>
\frac{p\rho}{1+\rho}-\epsilon'p\rho,\quad \textsl{for all
$\rho\in[0,1],$}\label{eq: pr0}
\end{equation}

Note that
\begin{eqnarray}
&&E_H\left[\exp\{-D {\tilde E}_o^{NF} (\frac{P|H|^2}{W_c},QPSK,\rho)\}\right] \nonumber\\
&=&E_H\left[  \left.e^{-D {\tilde E}_o^{NF}
(\frac{P|H|^2}{W_c},QPSK,\rho)}\right| |H|^2\le \frac{\xi
W_c}{P}\right] + E_H\left[  \left.e^{-D {\tilde E}_o^{NF}
(\frac{P|H|^2}{W_c},QPSK,\rho)}\right| |H|^2> \frac{\xi
W_c}{P}\right]
\nonumber\\
&\le& E_H\left[  \left.e^{-D (\frac{\rho}{1+\rho}-\epsilon'
\rho)\frac{P|H|^2}{W_c}}\right| |H|^2\le \frac{\xi W_c}{P}\right]
+
Pr(|H|^2> \frac{\xi W_c}{P})\label{eq: pr1}\\
&\le& E_H\left[  e^{- (\frac{\rho}{1+\rho}-\epsilon'
\rho)PT_c|H|^2}\right] + Pr(|H|^2> \frac{\xi W_c}{P}). \label{eq:
pr2}
\end{eqnarray}
The inequality in (\ref{eq: pr1}) comes from (\ref{eq: pr0}) and
the fact that $E_o(p,QPSK,\rho)\ge 0.$ For Rayleigh fading, we can
compute (\ref{eq: pr2}) and we have
$$
E_H\left[\exp\{-D {\tilde E}_o^{NF}
(\frac{P|H|^2}{W_c},QPSK,\rho)\}\right] \le
\frac{1}{1+(\frac{\rho}{1+\rho}-\epsilon' \rho)PT_c}
+e^{-\frac{\xi W_c}{P}}.
$$

We choose $\epsilon'$ such that $\epsilon'=\frac{\epsilon}{2P}.$
We can then find the corresponding $\xi$ with respect to this
choice of $\epsilon'.$ We then choose $W_c^{(\epsilon)}$ such that
$$
e^{-\frac{W_c^{(\epsilon)} \xi}{P}}<\frac{\epsilon}{2(1+P)}.
$$
It is straightforward to check that for all $W_c\ge
W_c^{(\epsilon)},$ (\ref{eq: pr5}) will be held and thus complete
the proof of this Lemma. \hfill $\diamond$

In summary, the first-order calculation gives us the following
theorem.
\begin{theorem}\label{thm: fading D}
Consider a coherent Rayleigh-fading channel (\ref{eq: vector
channel 1}), where $H$ is unit complex Gaussian random variable.
The sequence of input distributions of the channel is constrained
by ${\mathcal F}_{W_c}(P).$ Let $R(1/W_c)$ be the maximum rate at
which information can be transmitted on this channel, for a given
error-exponent constraint
$$
E(R,P,W_c)\ge z, \quad 0<z<z^*,
$$
where $z^*$ is defined by (\ref{eq: z star}). We have
\begin{equation}
R(0)=\lim_{W_c\rightarrow \infty} R(1/W_c) = \sup_{0\le\rho\le 1}
-\frac{z}{\rho}+\frac{1}{T_c}\frac{\ln\left(1+\frac{\rho
PT_c}{1+\rho}\right)}{\rho}.\label{eq: D 00}
\end{equation}
\hfill $\diamond$
\end{theorem}

Next we present a sufficient condition for a sequence of input
distributions $q_{W_c}({\bf x})$ to be first order optimal.

\begin{lemma}\label{lem: first order condition}
Assuming $0<z<z^*$, where $z^*$ is defined by (\ref{eq: z star}),
a sufficient condition for $\{q_{W_c}\}$ to be first-order optimal
is that
\begin{equation}
\lim_{W_c\rightarrow \infty} {\tilde
E}_o(P,q_{W_c},\rho^*,W_c)=\frac{1}{T_c}\ln(1+\frac{\rho^*
PT_c}{1+\rho^*}),\label{eq: first condition}
\end{equation}
where $\rho^*$ is the optimizing $\rho$ for (\ref{eq: D 00}).
\end{lemma}
Proof: Similar to the proof of Lemma~\ref{lem: first order
condition 1 0}.\hfill $\diamond$

Similar to the AWGN channel, in the fading channel with large
coherence bandwidth $W_c,$ it does not take much to be first-order
optimal. We restrict ourselves to those vector input distributions
which are i.i.d. in each dimension. We have the following lemma.

\begin{lemma}\label{lem: sufficient}
For i.i.d. input distributions, such that $q_{W_c}({\bf
x})=\Pi_{d=1}^D q(x_d),$ a sufficient condition for
$\{q_{W_c}({\bf x})\}\in {\mathcal F}_{W_c}(P)$ to be first-order
optimal is that $q(x)$ is symmetric around zero, i.e.
$$
q(x)=q(-x).
$$
\end{lemma}
{\bf Proof:} See Appendix~\ref{sec: sufficient fading}. \hfill
$\diamond$

\subsection{$\dot{R}(0)$ and second-order optimal condition}

To compute $\dot{R}(0),$ we first establish a relationship between
$\dot{R}(0)$ and the derivative of $E_o(P,\rho,W_c)$ with respect
to $1/W_c.$

\begin{theorem}\label{thm: rD 1}
If as $W_c$ goes to infinity, for each $\rho\in[0,1],$ the limit
of of $W_c\left[E_o(P,\rho,W_c)-E_o(P,\rho,\infty)\right]$ exists,
which we denote as $\dot{E}_o(P,\rho,\infty)$ and is a continuous
function in $\rho,$ and further,
\begin{equation}
W_c\left[E_o(P,\rho,W_c)-E_o(P,\rho,\infty)\right]\rightarrow
\dot{E}_o(P,\rho,\infty) \quad \textsl{uniformly for all
$\rho\in[0,1],$}\label{eq: pr10}
\end{equation}
$\dot{R}(0)$ can be determined as
\begin{equation}
\dot{R}(0)=\frac{\dot{E}_o(P,\rho^*,\infty)}{\rho^*}, \label{eq:
rD dot 0}
\end{equation}
where $\rho^*$ is the optimizing $\rho$ in (\ref{eq: D 00}).
\end{theorem}
Proof: The uniform convergence in (\ref{eq: pr10}) tells us: for
any $\epsilon>0,$ we can find $W_c^{(\epsilon)},$ such that for
all $W_c\ge W_c^{(\epsilon)},$ we have
\begin{equation}
\left| W_c\left[ E_o(P,\rho,W_c)-E_o(P,\rho,\infty)\right]
-\dot{E}_o(P,\rho,\infty)\right|\le \epsilon, \quad \forall \rho
\in[0,1].\label{eq: pr11}
\end{equation}
In other words, we know
$$
E_o(P,\rho,W_c)\le E_o(P,\rho,\infty)+
\frac{1}{W_c}\dot{E}_o(P,\rho,\infty) +\frac{\epsilon}{W_c}, \quad
\forall \rho.
$$
Applying Corollary~\ref{cor: equivalent form} here, we know that
for $W_c\ge W_c^{(\epsilon)},$
\begin{eqnarray*}
R &=& \sup_{\frac{z}{P}\le\rho \le 1} -\frac{z}{\rho}
+\frac{E_o(P,\rho,W_c)}{\rho}\\
&\le& \sup_{\frac{z}{P}\le\rho \le 1} -\frac{z}{\rho}
+\frac{E_o(P,\rho,\infty)+ \frac{1}{W_c}\dot{E}_o(P,\rho,\infty)
+\frac{\epsilon}{W_c}}{\rho}\\
&\le& \sup_{\frac{z}{P}\le\rho \le 1} -\frac{z}{\rho}
+\frac{E_o(P,\rho,\infty)}{\rho}+
\frac{\dot{E}_o(P,\rho,\infty)}{\rho W_c} +\frac{\epsilon P}{W_c
z}.
\end{eqnarray*}
Assume $\rho(W_c)$ is the optimizing $\rho$ for
$\sup_{\frac{z}{P}\le\rho \le 1} -\frac{z}{\rho}
+\frac{E_o(P,\rho,\infty)}{\rho}+
\frac{\dot{E}_o(P,\rho,\infty)}{\rho W_c}.$ Since the optimization
is over a compact interval, if $\frac{E_o(P,\rho,\infty)}{\rho}+
\frac{\dot{E}_o(P,\rho,\infty)}{\rho W_c}$ is continuous in
$\rho,$ the optimizing $\rho$ must exist. However, the first-order
calculation already gave us
$$
E_o(P,\rho,\infty)=\frac{1}{T_c}\ln(1+\frac{\rho PT_c}{1+\rho}),$$
which is a continuous function of $\rho,$ and we are assuming here
$\dot{E}_o(P,\rho,\infty)$ is continuous in $\rho,$ we must have
$\frac{E_o(P,\rho,\infty)}{\rho}+
\frac{\dot{E}_o(P,\rho,\infty)}{\rho W_c}$ continuous in $\rho$ as
well. Thus, it is well justified to denote $\rho(W_c)$ as the
optimizing $\rho$ here. Using this notation, we can further bound
$R(1/W_c)$ as follows
\begin{eqnarray*}
R(1/W_c) &\le& \left\{\sup_{\frac{z}{P}\le\rho \le 1}
-\frac{z}{\rho} +\frac{E_o(P,\rho,\infty)}{\rho}\right\}+
\frac{\dot{E}_o(P,\rho(W_c),\infty)}{\rho(W_c) W_c}
+\frac{\epsilon
P}{W_cz} \\
&=& R(0)+ \frac{\dot{E}_o(P,\rho(W_c),\infty)}{\rho(W_c) W_c}
+\frac{\epsilon P}{W_c z}.
\end{eqnarray*}
If we define $\dot{\overline{R}}(0)=\limsup_{W_c\rightarrow
\infty} W_c[R(1/W_c)-R(0)],$ we have
\begin{eqnarray*}
\dot{\overline{R}}(0) &\le& \limsup_{W_c\rightarrow \infty}
\frac{\dot{E}_o(P,\rho(W_c),\infty)}{\rho(W_c)}+\frac{\epsilon
P}{z}\\
&=& \frac{\dot{E}_o(P,\rho^*,\infty)}{\rho^*} +\frac{\epsilon
P}{z}.
\end{eqnarray*}
Here we use the fact
\begin{equation}
\lim_{W_c\rightarrow \infty} \rho(W_c)\rightarrow \rho^*
\label{eq: rho}
\end{equation}
and the assumption that $\dot{E}_o(P,\rho,\infty)$ is a continuous
function in $\rho.$ The proof of (\ref{eq: rho}) is similar to
Appendix~\ref{sec: limit}.

Letting $\epsilon$ goes to $0,$ we know
$$
\dot{\overline{R}}(0) \le
\frac{\dot{E}_o(P,\rho^*,\infty)}{\rho^*}.
$$

On the other hand, (\ref{eq: pr11}) also implies
\begin{eqnarray*}
R(1/W_c) &\ge& \sup_{\frac{z}{P}\le\rho \le 1} -\frac{z}{\rho}
+\frac{E_o(P,\rho,\infty)+ \frac{1}{W_c}\dot{E}_o(P,\rho,\infty)
-\frac{\epsilon}{W_c}}{\rho}\\
&\ge& \sup_{\frac{z}{P}\le\rho \le 1} -\frac{z}{\rho}
+\frac{E_o(P,\rho,\infty)}{\rho}+
\frac{\dot{E}_o(P,\rho,\infty)}{\rho W_c} -\frac{\epsilon P}{W_c z} \\
&\ge& -\frac{z}{\rho^*} +\frac{E_o(P,\rho^*,\infty)}{\rho^*}+
\frac{\dot{E}_o(P,\rho^*,\infty)}{\rho^* W_c} -\frac{\epsilon
P}{W_c z}
\\
&=& R(0) + \frac{\dot{E}_o(P,\rho^*,\infty)}{\rho^* W_c}
-\frac{\epsilon P}{W_c z}.
\end{eqnarray*}
Letting $\epsilon\rightarrow 0,$ we have
$$
\dot{\underline{R}}(0)=\liminf_{W_c\rightarrow \infty}
W_c[R(1/W_c)-R(0)] \ge \frac{\dot{E}_o(P,\rho^*,\infty)}{\rho^*}.
$$
\hfill $\diamond$

Next we verify the uniform convergence assumption needed in
Theorem~\ref{thm: rD 1}.
\begin{lemma}\label{lem: uni2}
As $W_c$ goes to infinity, we have
\begin{equation}
W_c\left[E_o(P,\rho,W_c)-E_o(P,\rho,\infty)\right] \rightarrow
-\frac{\rho P^2}{(1+\rho)(1+\rho+ {\rho PT_c})^2}\quad
\textsl{uniformly for $\rho\in[0,1].$}
\end{equation}
\end{lemma}
Proof: To show the uniform convergence result, we find both an
upper bound and a lower bound for
$$
W_c\left[E_o(P,\rho,W_c)-E_o(P,\rho,\infty)\right]
$$
and both bounds converges uniformly to $-\frac{\rho
P^2}{(1+\rho)(1+\rho+ \rho PT_c)^2}.$

For notational convenience, we introduce the notation
$\delta(\frac{1}{W_c^m})$ which indicates a term satisfying
$$
\lim_{W_c\rightarrow \infty} W_c^m \delta(\frac{1}{W_c^m})=0,\quad
\textsl{uniformly for $\rho\in[0,1].$}
$$
Using this notation, what we need to show here is
\begin{eqnarray}
E_o(P,\rho,W_c)&\le& \frac{1}{T_c}\ln(1+\frac{\rho
PT_c}{1+\rho})-\frac{\rho
P^2}{W_c(1+\rho)(1+\rho+ {\rho P}T_c)^2} +\delta(\frac{1}{W_c});\label{eq: up}\\
E_o(P,\rho,W_c)&\ge& \frac{1}{T_c}\ln(1+\frac{\rho
PT_c}{1+\rho})-\frac{\rho P^2}{W_c(1+\rho)(1+\rho+ {\rho P}T_c)^2}
+\delta(\frac{1}{W_c})\label{eq: down}.
\end{eqnarray}

For the upper bound, we again use the inequality (\ref{eq: a00}),
which gives us
$$
E_o(P,\rho,W_c)\le -\frac{1}{T_c}\ln
E_H[e^{-DE_o^{NF}(\frac{P|H|^2}{W_c},\rho)}].
$$

We showed that
$\frac{\frac{E_o^{NF}(p,\rho)}{p\rho}-\frac{1}{1+\rho}}{p}$
converges to $-\frac{1}{2(1+\rho)^3}$ uniformly, or equivalently
saying, for any $\epsilon>0,$ we can find $\xi>0,$ such that for
any $p\le \xi,$
$$
\frac{\rho p}{1+\rho} -\frac{\rho p^2}{2(1+\rho)^3} -\epsilon \rho
p^2\le E_o^{NF}(p,\rho)\le \frac{\rho p}{1+\rho} -\frac{\rho
p^2}{2(1+\rho)^3} +\epsilon \rho p^2.
$$
Thus, we have
\begin{eqnarray*}
&&E_H[e^{-DE_o^{NF}(\frac{P|H|^2}{W_c},\rho)}] \\
&\ge& E_H\left[\left.
e^{-DE_o^{NF}(\frac{P|H|^2}{W_c},\rho)}\right|
|H|^2\le \frac{W_c\xi}{P}\right]\\
&\ge& E_H\left[\left. e^{-D(\frac{\rho P|H|^2}{W_c(1+\rho)}
-\frac{\rho P^2|H|^4}{2W_c^2(1+\rho)^3} +\frac{\epsilon\rho
P^2|H|^4}{W_c^2})}\right| |H|^2\le \frac{W_c\xi}{P}\right]\\
&\ge& E_H\left[\left. e^{-\frac{\rho PT_c|H|^2}{1+\rho}} \left(1+
\frac{\rho P^2|H|^4T_c}{2W_c(1+\rho)^3} -\frac{\epsilon\rho
P^2|H|^4T_c}{W_c}\right)\right| |H|^2\le \frac{W_c\xi}{P}\right]\\
&=& E_H\left[e^{-\frac{\rho PT_c|H|^2}{1+\rho}} \left(1+
\frac{\rho P^2|H|^4T_c}{2W_c(1+\rho)^3} -\frac{\epsilon\rho
P^2|H|^4T_c}{W_c}\right)\right]\\
&&- E_H\left[\left. e^{-\frac{\rho PT_c|H|^2}{1+\rho}} \left(1+
\frac{\rho P^2|H|^4T_c}{2W_c(1+\rho)^3} -\frac{\epsilon\rho
P^2|H|^4T_c}{W_c}\right)\right| |H|^2\ge \frac{W_c\xi}{P}\right]\\
&\ge& E_H\left[e^{-\frac{\rho PT_c|H|^2}{1+\rho}} \left(1+
\frac{\rho P^2|H|^4T_c}{2W_c(1+\rho)^3} -\frac{\epsilon\rho
P^2|H|^4T_c}{W_c}\right)\right]\\
&&- e^{-\frac{\rho D\xi}{1+\rho}}E_H\left[\left(1+ \frac{\rho
P^2|H|^4T_c}{2W_c(1+\rho)^3} -\frac{\epsilon\rho
P^2|H|^4T_c}{W_c}\right)\right]\\
&=& \frac{1}{1+\frac{\rho
PT_c}{1+\rho}}+\frac{\left(\frac{\rho}{(1+\rho)^3} -2\epsilon
\rho\right) \frac{P^2T_c}{W_c}}{\left(1+\frac{\rho
PT_c}{1+\rho}\right)^3} - e^{-\frac{\rho D\xi}{1+\rho}}\left(1+
\frac{\rho P^2T_c}{W_c(1+\rho)^3} -\frac{2\epsilon\rho
P^2T_c}{W_c}\right).
\end{eqnarray*}

Thus,
\begin{eqnarray*}
&&E_o(P,\rho,W_c)\\
&\le& -\frac{1}{T_c}\ln\left\{\frac{1}{1+\frac{\rho
PT_c}{1+\rho}}+\frac{\left(\frac{\rho}{(1+\rho)^3} -2\epsilon
\rho\right) \frac{P^2T_c}{W_c}}{\left(1+\frac{\rho
PT_c}{1+\rho}\right)^3} - e^{-\frac{\rho D\xi}{1+\rho}}\left(1+
\frac{\rho P^2T_c}{W_c(1+\rho)^3} -\frac{2\epsilon\rho
P^2T_c}{W_c}\right)\right\} \\
&=& \frac{1}{T_c}\ln(1+\frac{\rho PT_c}{1+\rho})\\
&&-\frac{1}{T_c}\ln\left\{1 +\frac{\left(\frac{\rho}{(1+\rho)^2}
-2\epsilon \rho\right) \frac{P^2T_c}{W_c}}{\left(1+\frac{\rho
PT_c}{1+\rho}\right)^2} -(1+\frac{\rho PT_c}{1+\rho})
e^{-\frac{\rho D\xi}{1+\rho}}\left(1+ \frac{\rho
P^2T_c}{W_c(1+\rho)^3} -\frac{2\epsilon\rho
P^2T_c}{W_c}\right)\right\}\\
&=& \frac{1}{T_c}\ln(1+\frac{\rho PT_c}{1+\rho})
-\frac{1}{T_c}\ln\left\{1 + \frac{\rho
P^2T_c}{(1+\rho)(1+\rho+\rho PT_c)^2W_c} -\frac{2\epsilon
\rho P^2T_c}{(1+\frac{\rho PT_c}{1+\rho})^2W_c}\right. \\
&&\left. -(1+\frac{\rho PT_c}{1+\rho}) e^{-\frac{\rho
D\xi}{1+\rho}}\left(1+ \frac{\rho P^2T_c}{W_c(1+\rho)^3}
-\frac{2\epsilon\rho P^2T_c}{W_c}\right)\right\}.
\end{eqnarray*}

Since we can choose an arbitrary small $\epsilon$ here, it is
straightforward to show that the term
$$
\frac{2\epsilon \rho P^2T_c}{(1+\frac{\rho PT_c}{1+\rho})^2W_c}
-(1+\frac{\rho PT_c}{1+\rho}) e^{-\frac{\rho
D\xi}{1+\rho}}\left(1+ \frac{\rho P^2T_c}{W_c(1+\rho)^3}
-\frac{2\epsilon\rho P^2T_c}{W_c}\right)
$$
is actually $\delta(\frac{1}{W_c}).$ Thus, we have
\begin{eqnarray*}
&&E_o(P,\rho,W_c)\\
&\le& \frac{1}{T_c}\ln(1+\frac{\rho
PT_c}{1+\rho})-\frac{1}{T_c}\ln\left(1 + \frac{\rho
P^2T_c}{(1+\rho)(1+\rho+\rho PT_c)^2W_c} +\delta(\frac{1}{W_c})\right)\\
&=& \frac{1}{T_c}\ln(1+\frac{\rho PT_c}{1+\rho})-\frac{\rho
P^2}{(1+\rho)(1+\rho+\rho PT_c)^2W_c} +\delta(\frac{1}{W_c}).
\end{eqnarray*}

For the lower bound, we again use the QPSK calculation:
$$
E_o(P,\rho,W_c)\ge {\tilde E}_o(P,QPSK,\rho,W_c)=-\frac{1}{T_c}\ln
E_H[\exp\{-D {\tilde E}_o^{NF}(\frac{P|H|^2}{W_c},QPSK,\rho)\}].
$$

In last section, we have already shown that
$$
\frac{\frac{{\tilde E}_o^{NF}(p,QPSK,\rho)}{\rho
p}-\frac{1}{1+\rho}}{p} \rightarrow -\frac{1}{2(1+\rho)^3},
\textsl{uniformly for $\rho\in[0,1].$}
$$
Equivalently, for any $\epsilon>0,$ we can find $\xi>0$ such that
for all $\rho\in[0,1],$ and all $p<\xi,$
$$
\frac{\rho p}{1+\rho} -\frac{\rho p^2}{2(1+\rho)^3} -\epsilon \rho
p^2\le {\tilde E}_o^{NF}(p,QPSK,\rho)\le \frac{\rho p}{1+\rho}
-\frac{\rho p^2}{2(1+\rho)^3} +\epsilon \rho p^2.
$$

Thus, we have
\begin{eqnarray}
&&E_H[e^{-D {\tilde E}_o^{NF}(\frac{P|H|^2}{W_c},QPSK,\rho)}] \nonumber\\
&=& E_H[\left. e^{-D {\tilde
E}_o^{NF}(\frac{P|H|^2}{W_c},QPSK,\rho)}\right| |H|^2\le
\frac{W_c\xi}{P}]+ E_H[\left. e^{-D {\tilde
E}_o^{NF}(\frac{P|H|^2}{W_c},QPSK,\rho)}\right| |H|^2\ge
\frac{W_c\xi}{P}] \nonumber\\
&\le& E_H[\left. e^{-D \left(\frac{\rho P|H|^2}{W_c(1+\rho)}
-\frac{\rho P^2 |H|^4}{2W_c^2(1+\rho)^3} -\epsilon \rho
\frac{P^2|H|^4}{W_c^2}\right)}\right| |H|^2\le \frac{W_c\xi}{P}]+
e^{-\frac{W_c\xi}{P}}\nonumber\\
&=& E_H[\left. e^{-\frac{\rho P|H|^2T_c}{1+\rho} +\frac{\rho P^2
|H|^4T_c}{2W_c (1+\rho)^3} +\epsilon \rho
\frac{P^2|H|^4T_c}{W_c}}\right| |H|^2\le \frac{W_c\xi}{P}]+
e^{-\frac{W_c\xi}{P}}\label{eq: pr6}.
\end{eqnarray}

A useful inequality we can use here is the following
\begin{equation}
e^{t}\le 1+t+t^2 e^t \quad \forall t\in R.\label{eq: inequation}
\end{equation}
To show the validity of (\ref{eq: inequation}), we check (\ref{eq:
inequation}) for two cases: $t\ge 1$ and $t<1.$ When $t\ge 1,$
(\ref{eq: inequation}) is trivial. When $t<1,$ we start with the
following well-known inequality: $e^{-t}\ge 1-t.$ Since $t<1,$
this leads to
$$
e^{t}\le \frac{1}{1-t}=\frac{1+t}{1-t^2}.
$$
From here, it is easy to see that (\ref{eq: inequation}) is true.

Define
$$
\eta=\frac{\rho }{2(1+\rho)^3}+ \epsilon \rho.
$$
Applying (\ref{eq: inequation}) in (\ref{eq: pr6}), we have
\begin{eqnarray}
&&E_H[e^{-D {\tilde E}_o^{NF}(\frac{P|H|^2}{W_c},QPSK,\rho)}]\nonumber\\
&\le& E_H[\left. e^{-\frac{\rho P|H|^2T_c}{1+\rho}} \left(1+\eta
\frac{P^2 |H|^4T_c}{W_c}+\eta^2 \frac{P^4 |H|^8T_c^2}{W_c^2}
e^{\eta \frac{P^2 |H|^4T_c}{W_c}} \right)\right| |H|^2\le
\frac{W_c\xi}{P}]+
e^{-\frac{W_c\xi}{P}}\nonumber\\
&\le& E_H[e^{-\frac{\rho P|H|^2T_c}{1+\rho}} \left(1+\eta
\frac{P^2 |H|^4T_c}{W_c}\right)]+ E_H[\left. \eta^2 \frac{P^4
|H|^8T_c^2}{W_c^2} e^{-\frac{\rho P|H|^2T_c}{1+\rho} +\eta
\frac{P^2 |H|^4T_c}{W_c}} \right|
|H|^2\le \frac{W_c\xi}{P}]\nonumber\\
&&+ e^{-\frac{W_c\xi}{P}}.\label{eq: pr7}
\end{eqnarray}
For the second term in (\ref{eq: pr7}), since $|H|^2\le
\frac{W_c\xi}{P},$ we have
\begin{eqnarray*}
-\frac{\rho P|H|^2T_c}{1+\rho} +\eta \frac{P^2 |H|^4T_c}{W_c}
&\le&
-\frac{\rho P|H|^2T_c}{1+\rho} +\eta {P |H|^2 T_c\xi} \\
&=& -\frac{\rho PT_c|H|^2}{1+\rho} + \left(\frac{\rho
}{2(1+\rho)^3}+ \epsilon \rho \right) P T_c|H|^2 \xi.
\end{eqnarray*}
For sufficiently small $\epsilon$ and $\xi,$ (for example,
$\epsilon<1$ and $\xi<1,$) we have
$$
-\frac{\rho PT_c|H|^2}{1+\rho} +\eta \frac{P^2 |H|^4T_c}{W_c}\le
0.
$$
Thus, we can further bound (\ref{eq: pr7}) as follows:
\begin{eqnarray*}
&&E_H[e^{-D {\tilde E}_o^{NF}(\frac{P|H|^2}{W_c},QPSK,\rho)}]\nonumber\\
&\le& E_H[e^{-\frac{\rho P|H|^2T_c}{1+\rho}} \left(1+\eta
\frac{P^2 |H|^4T_c}{W_c}\right)]+ E_H[\left. \eta^2 \frac{P^4
|H|^8T_c^2}{W_c^2} \right|
|H|^2\le \frac{W_c\xi}{P}]+ e^{-\frac{W_c\xi}{P}} \\
&\le& \frac{1}{1+\frac{\rho
PT_c}{1+\rho}}+\frac{\left(\frac{\rho}{(1+\rho)^3} +2\epsilon
\rho\right) \frac{P^2T_c}{W_c}}{\left(1+\frac{\rho
PT_c}{1+\rho}\right)^3}
+E_H[\eta^2 \frac{P^4 |H|^8T_c^2}{W_c^2}]+ e^{-\frac{W_c\xi}{P}} \\
&=& \frac{1}{1+\frac{\rho
PT_c}{1+\rho}}+\frac{\frac{\rho}{(1+\rho)^3}
\frac{P^2T_c}{W_c}}{\left(1+\frac{\rho PT_c}{1+\rho}\right)^3}
+\delta(\frac{1}{W_c}).
\end{eqnarray*}
Thus,
\begin{eqnarray*}
&&E_o(P,\rho,W_c)\\
&\ge& \frac{1}{T_c}\ln(1+\frac{\rho PT_c}{1+\rho})
-\frac{1}{T_c}\ln\left(1+\frac{\rho P^2T_c}{(1+\rho)(1+\rho+\rho P
T_c)^2 W_c} +\delta(\frac{1}{W_c})\right)
\\
&=&\frac{1}{T_c}\ln(1+\frac{\rho P}{1+\rho}) -\frac{\rho
P^2}{(1+\rho)(1+\rho+\rho PT_c)^2 W_c} +\delta(\frac{1}{W_c}).
\end{eqnarray*}

Thus, we have shown both (\ref{eq: up}) and (\ref{eq: down}). From
these two equations, it is easy to see the uniform convergence as
claimed in Lemma~\ref{lem: uni2}. \hfill $\diamond$

Combining Lemma~\ref{lem: uni2} and Theorem~\ref{thm: rD 1}, we
have the following theorem.
\begin{theorem}\label{thm: fading D 2}
Consider a coherent Rayleigh-fading vector channel (\ref{eq:
vector channel 1}), where $H$ is a unit complex Gaussian random
variable. Let $R(1/W_c)$ be the maximum rate at which information
can be transmitted on this channel. The sequence of input
distributions of the channel is constrained by ${\mathcal
F}_{W_c}(P).$ For a given error-exponent constraint
$$
E(R,P,W_c)\ge z, \quad 0<z<z^*,
$$
where $z^*$ is defined by (\ref{eq: z star}), we have
\begin{equation}\label{eq: dot R}
\dot{R}(0)=-\frac{P^2}{(1+\rho^*)(1+\rho^*+ {\rho^* PT_c})^2},
\end{equation}
where $\rho^*$ is the optimizing $\rho$ in (\ref{eq: D 00}).
\hfill $\diamond$
\end{theorem}

\begin{theorem}\label{thm: BPSK and QPSK D}
Both BPSK and QPSK are first-order optimal for any given
$z\in(0,z^*)$; however, only QPSK is second-order optimal.
\end{theorem}
Proof: The first-order optimality of BPSK and QPSK can be easily
seen from Lemma~\ref{lem: sufficient}. In the proof of
Lemma~\ref{lem: uni2}, we essentially showed that by choosing the
input distribution of QPSK,
$$
{\tilde E}_o(P,QPSK,\rho,W_c)\ge \frac{1}{T_c}\ln(1+\frac{\rho
PT_c}{1+\rho}) - \frac{\rho P^2}{(1+\rho)(1+\rho+\rho PT_c)^2
W_c}+\delta(\frac{1}{W_c}).
$$
On the other hand, it was also shown in the proof of
Lemma~\ref{lem: uni2} that
$$
{\tilde E}_o(P,QPSK,\rho,W_c)\le E_o(P,\rho,W_c) \le
\frac{1}{T_c}\ln(1+\frac{\rho PT_c}{1+\rho}) - \frac{\rho
P^2}{(1+\rho)(1+\rho+\rho PT_c)^2 W_c}+\delta(\frac{1}{W_c}).
$$
Thus, we must have
$$
W_c[{\tilde E}_o(P,QPSK,\rho,W_c)-\frac{1}{T_c}\ln(1+\frac{\rho
PT_c}{1+\rho})]\rightarrow - \frac{\rho P^2}{(1+\rho)(1+\rho+\rho
PT_c)^2} \quad \textsl{uniformly for $\rho\in[0,1].$}
$$
Following a similar argument as in Theorem~\ref{thm: rD 1}, we can
easily obtain
$$
\dot{\tilde{R}}(0)=\dot{R}(0)= -\frac{P^2}{(1+\rho^*)(1+\rho^*+
{\rho^* PT_c})^2}.
$$
For BPSK, using the result in last section regarding BPSK, we can
obtain that
$$
W_c[{\tilde E}_o(P,QPSK,\rho,W_c)-\ln(1+\frac{\rho
P}{1+\rho})]\rightarrow - \frac{2\rho P^2}{(1+\rho)(1+\rho+\rho
PT_c)^2} \quad \textsl{uniformly for $\rho\in[0,1].$}
$$
Thus,
$$
\dot{\tilde{R}}(0)= -\frac{2P^2}{(1+\rho^*)(1+\rho^*+ {\rho^*
PT_c})^2}<\dot{R}(0).
$$
Therefore, QPSK is near optimal while BPSK is not.  \hfill
$\diamond$

\section{Conclusions}\label{sec: conclusions}

In this paper, we have studied the maximum rate at which
information transmission is possible in additive Gaussian noise
channels and coherent fading channels, for a given error exponent
in the wideband regime. Given a desired error exponent, our main
contribution is the calculation of the above rate and its
derivative in the limit when the available bandwidth goes to
$\infty.$ For fading channels, we focus on the case when the
coherence bandwidth $W_c$ is large. This also leads to a notion of
near-optimality of input distributions, where a sequence of
distributions is defined to be near-optimal if it achieves both
the rate and its derivative in the infinite bandwidth limit. As in
\cite{ver02}, we show that for both AWGN and coherent fading
channels, while QPSK is near-optimal, BPSK is not.

This result is surprising to some extent. Generally, it is not
well-understood as to what signaling scheme is optimal, i.e.,
given a coding rate, it is difficult to find the input
distribution that gives the smallest probability of decoding
error. In this paper, we consider the problem from an alternate
point of view, we fix a given error exponent, and consider optimal
signaling schemes that gives the largest communication rate. The
capacity-achieving schemes, which corresponds to zero error
exponent, are not necessarily the best schemes from the error
exponent point of view. However, the results in this paper tell
us, in the wideband regime, QPSK is near-optimal with respect to a
nonzero error exponent just as it is near-optimal for the capacity
case for both AWGN and coherent fading channels. Thus, it can not
only achieves capacity, but also achieves the the best probability
of decoding error, in the wideband regime.

\appendix

\section{The reliability function}\label{sec: reliability function}

In this section, we will summarize some important bounds on the
{\sl reliability function}. To be consistent with other
literature, we will use the traditional notation for the
reliability function (as just a function of $R$) to present the
bounds. Please note that elsewhere in this paper, the reliability
function is defined as in (\ref{eq: def E}).

\begin{define}\cite{gal68}
Let $P_e(N,R)$ be the minimum probability of error for any block
code of block length $N$ and rate $R$ for a given channel. The
reliability function $E(R)$ of this channel is defined as

\begin{equation}\label{eq: reliability function}
E(R)=\lim_{N\rightarrow \infty} -\frac{\ln P_e(N,R)}{N}.\label{eq:
tee}
\end{equation}
\end{define}\hfill $\diamond$

In \cite{gal65,gal68}, Gallager provides an upper bound for the
probability of error of discrete memoryless channel (DMC). This
result can be extended to a discrete-time memoryless channel with
a continuous alphabet associated with an average power constraint,
as stated in Theorem 10 of \cite{gal65}.

\begin{theorem}\label{thm: random coding}
\cite{gal65,gal68} Let $f(y|x)$ be the transition probability
density of a discrete-time memoryless channel and assume that each
codeword is constrained to satisfy $\sum_{n=1}^N |x_n|^2\le NP$.
Then, for any block code with length $N$ and rate $R$, there
exists a code for which
\begin{equation}
E(R)\ge E_r(R), \label{eq: random coding bound}
\end{equation}
with
\begin{eqnarray}
&E_r(R)&=\sup_{0\le \rho\le 1} -\rho R+E_o(\rho)\nonumber\\
&E_o(\rho)&=\sup_{E_{x}(|x|^2)\le P}\sup_{\beta\ge 0}
-\ln\int\left(\int
q(x)e^{\beta\left(|x|^2-P\right)}f(y|x)^{\frac{1}{1+\rho}}
d{x}\right)^{1+\rho} d{y}. \label{eq: Enote0}
\end{eqnarray}
\hfill $\diamond$
\end{theorem}
We will refer to $E_r(R)$ as the {\sl random-coding exponent} of
the channel and $\beta$ as the power-constraint parameter.

To find a lower bound on the error probability (or equivalently,
an upper bound on the reliability function) for a given channel is
a much harder problem. In \cite{fan61}, Fano derived the
sphere-packing lower bound for a discrete-memoryless channel (DMC)
in a heuristic manner. The first rigorous proof was provided by
Shannon et. al. in \cite{shagalber67}. In \cite{bla87}, a more
intuitive and simpler proof was provided by Blahut by connecting
the decoding error probability to a binary hypothesis-testing
problem. The sphere-packing exponent $E_{sp}(R)$ coincides with
the random-coding exponent $E_r(R)$ for a rate larger than a
critical rate $R_{crit}$, when the optimizing $\rho$ equals to
$1$. Gallager also extended the lower bound result to a DMC with
power constraint in \cite{gal68} and noted that the random-coding
exponent in this case also coincides with the sphere-packing
exponent for $R>R_{crit}$. In a later work \cite{gal87}, he
indicates that the lower bound is also applicable to a
discrete-time, continuous channel with a finite, discrete set of
input symbols and continuous output alphabet.

\begin{theorem}\label{thm: sphere packing}
Consider a discrete-time memoryless channel with a discrete finite
input alphabet $\{{x}_1,{x}_2,\cdots,{x}_K\}$ and the average
input power is constrained by $P.$ Let $f(y|x)$ be the transition
probability distribution. For any $(N,R)$ code, we have
\begin{eqnarray}
E(R)\le E_{sp}(R),
\end{eqnarray}
with
\begin{eqnarray}
E_{sp}(R)&=&\sup_{\rho \ge 0} -\rho R+E_o(\rho),\nonumber\\
E_o(\rho)&=&\sup_{E_{x}(\|{x}\|^2)\le P}\sup_{\beta\ge 0} -\ln
\int\left(\sum_{k=1}^K
q({x_k})e^{\beta\left(|{x_k}|^2-P\right)}f(y|x_k)^{\frac{1}{1+\rho}}\right)^{1+\rho}
d{y}. \label{eq: Enote00}
\end{eqnarray}
\hfill $\diamond$
\end{theorem}

As in \cite{gal68}, using the Kuhn-Tucker conditions, we can
derive a necessary and sufficient condition for $q$ and $\beta$ to
be optimal.

\begin{lemma}\cite{gal68}\label{lem: kuhn-tucker}
A necessary and sufficient condition for $q$ and $\beta$ to
optimize (\ref{eq: Enote00}) is
\begin{equation}
\int
\alpha({y})^{\rho}e^{\beta(|{x_k}|^2-P)}f({y}|{x_k})^{\frac{1}{1+\rho}}
d{y}\ge \int \alpha({y})^{1+\rho} d{y}, \quad \forall
x_k\label{eq: kuhn tucker}
\end{equation}
with equality if $q({x_k})>0,$ where
\begin{equation}
\alpha({y})=\sum_{k=1}^K
q({x_k})e^{\beta(|{x_k}|^2-P)}f({y}|{x_k})^{\frac{1}{1+\rho}}.
\label{eq: alpha x}
\end{equation}
\end{lemma}

Unfortunately, the sphere-packing result can not be applied to the
case with an infinite number of input symbols. Thus, throughout
this paper, we only consider input distributions with discrete and
finite input alphabet. If we constrain the input distributions to
be in $D(P)$ as defined by Definition~\ref{def: Dp}, it is easy to
see that the only difference between the random-coding exponent
and the sphere-packing exponent is the range of $\rho$ on which
the optimization is performed. Thus, for $R$ larger than the
critical rate $R_{crit}$, where the optimizing $\rho=1,$ the
random-coding exponent and sphere-packing exponent coincide with
each other and give the true expression for the reliability
function.

\section{Proof of Lemma~\ref{lem: quarter}}\label{sec: quarter}

\begin{figure}
\center \epsfig{file=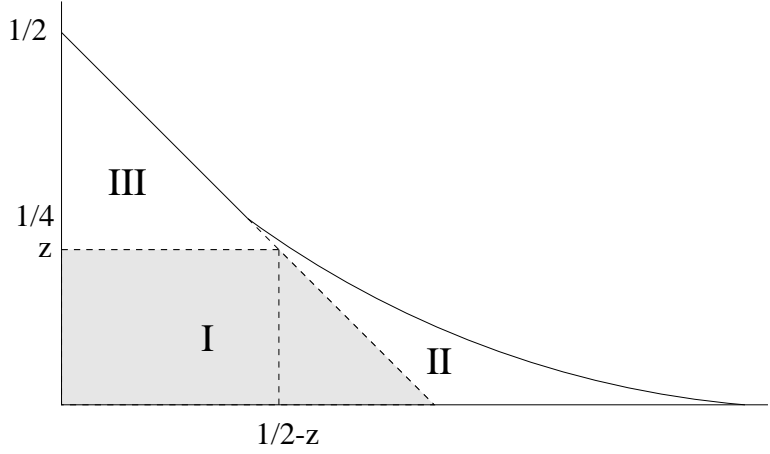,width=4.0in} \caption{The
reliability function for AWGN channel with infinite
bandwidth}\label{fig: quarter}
\end{figure}

We prove this lemma by contradiction. Given an error exponent
constraint $z<\frac{1}{4},$ assume that for any $B_z<\infty,$ we
can find $B\ge B_z,$ such that $R(1/B)\ne R_r(1/B).$ A direct
consequence of this assumption is that we know the critical rate
at bandwidth $B$, which we denote as $R_{crit}(1/B),$ satisfies
\begin{equation}
E(R_{crit}(1/B))< z.\label{eq: contra}
\end{equation}
For simplicity, in this proof, we assume $P=1.$ The infinite
bandwidth reliability function of the AWGN channel is shown in
Figure~\ref{fig: quarter}. Now we study the possible position of
the point $(R_{crit}(1/B),z_{crit}(1/B))$ in this figure.

Since the error exponent for any given rate is a non-decreasing
function of $B,$ a trivial observation we can make right away is
that the tuple $(R_{crit}(1/B),z_{crit}(1/B))$ has to be below the
infinite bandwidth reliability function. Equation (\ref{eq:
contra}) further tells us that it can not be in region III. Now we
argue that $(R_{crit}(1/B),z_{crit}(1/B))$ can not be in region II
either. If the tuple is in region II, we know the linear part of
the random-coding exponent will intersect the infinite-bandwidth
reliability function curve and thus for some communication rate,
using a finite bandwidth $B/2$ is than using infinite bandwidth.
This cannot be true and as a consequence,
$(R_{crit}(1/B),z_{crit}(1/B))$ can only be in region I, which is
the shaded region.

Next consider the random-coding exponent for rate $1/2-z.$ It is
straightforward to see that
$$
E_r(1/2-z,B)<z<E_r(1/2-z,\infty).
$$
Combining this with our assumption, we know that the following
equation can not be true:
$$
\lim_{B\rightarrow \infty} E_r(1/2-z,B)=E_r(1/2-z,\infty).
$$

However, it is well known that for any rate between $0$ and
capacity, the random-coding exponent converges to the
infinite-bandwidth error exponent as the bandwidth increases to
infinity. Thus, we have a contradiction.

\section{Proof of Theorem~\ref{thm: equivalent form}}\label{sec:
equivalent form}

The error-exponent constraint gives us
$$
pz=\sup_{0\le\rho\le 1} -\rho r+E_o(p,\rho),
$$
which is equivalent to say the following
\begin{itemize}
\item[1] For any $\rho\in[0,1],$ we always have
\begin{equation}
pz\ge -\rho r + E_o(p,\rho).\label{eq: y1}
\end{equation}
\item[2] For any $\epsilon>0,$ we can find $\rho_\epsilon,$ such that
 \begin{equation}
pz-\epsilon\le -\rho_\epsilon r+ E_o(p,\rho_\epsilon).\label{eq: y2}
\end{equation}
\end{itemize}

Similarly, what we want show is equivalent to the following
\begin{itemize}
\item[1] For any $\rho\in[0,1],$ we always have
\begin{equation}
r\ge -\frac{pz}{\rho} +\frac{ E_o(p,\rho)}{\rho}.\label{eq: y3}
\end{equation}
\item[2] For any $\eta>0,$ we can find $\rho_\eta,$ such that
 \begin{equation}
r-\eta\le -\frac{pz}{\rho_\eta} +\frac{
  E_o(p,\rho_\eta)}{\rho_\eta}.\label{eq: y4}
\end{equation}
\end{itemize}

It is easy to see that (\ref{eq: y3}) follows directly from (\ref{eq:
  y1}). Thus, it suffices to show (\ref{eq: y4}) is true. To do this,
  first we construct an $\epsilon$ from $\eta$ as follows
\begin{equation}
\epsilon=\frac{pz\eta}{p-r+\eta}.\label{eq: y5}
\end{equation}
First we check that $\epsilon>0.$ This is true if we have $p>r.$ Note
from the coding theorem, we know the largest rate available for
reliable communication, which is defined as capacity, is equal to
$\log(1+p)$ (nats per symbol) for AWGN channel. Hence, $r\le
c=log(1+p)\le p.$

From (\ref{eq: y2}), we know we could find a $\rho_\epsilon\in[0,1]$
such that
$$
r\le -\frac{pz}{\rho_\epsilon} +\frac{
  E_o(p,\rho_\epsilon)}{\rho_\epsilon}+\frac{\epsilon}{\rho_\epsilon}.
$$

Next we show $\frac{\epsilon}{\rho_\epsilon}\le \eta.$

From Lemma~\ref{lem: upper bound}, we know from (\ref{eq: y2} that
$$
pz-\epsilon\le \rho_\epsilon r+\frac{p\rho_\epsilon}{1+\rho_\epsilon}
\le -\rho_\epsilon r+p \rho_\epsilon =(p-r)\rho_\epsilon.
$$
Hence, we must have
$$
\rho_\epsilon\ge \frac{pz-\epsilon}{p-r}.
$$
Thus,
$$
\frac{\epsilon}{\rho_\epsilon}\le
\frac{\epsilon(p-r)}{pz-\epsilon}.
$$
Use (\ref{eq: y5}) to get
$$
\frac{\epsilon}{\rho_\epsilon}\le \eta.
$$

In other words, for any $\eta>0,$ we simply use
$\rho_\eta=\rho_\epsilon,$ and we will have (\ref{eq: y4}), which
completes the proof of this theorem.

\section{Proof of $\lim_{p\rightarrow 0}
\rho(p)=\rho^*$}\label{sec: limit}

We need to show that
$$
\lim_{p\rightarrow 0} \rho(p)=\rho^*,
$$
where $\rho(p)$ is the optimizing $\rho$ for the following
equation
$$
\rho(p)=\arg\sup_{0\le\rho\le 1}
-\frac{z}{\rho}+\frac{\dot{E}_o(0,\rho)}{\rho}+\frac{p\ddot{E}_o(0,\rho)}{2\rho},
$$
and $\rho^*$ is defined as follows
$$
\rho^*=\arg\sup_{0\le\rho\le 1} -\frac{z}{\rho}+\frac{1}{1+\rho}
=\frac{\sqrt{z}}{1-\sqrt{z}}.
$$
The assumption we can use here is that
$\frac{\ddot{E}_o(0,\rho)}{\rho}$ is a continuous and bounded
function in $\rho$ for $\rho\in[0,1].$ A direct consequence of
this assumption is that as $p\rightarrow 0,$
\begin{equation}
\frac{\dot{E}_o(0,\rho)}{\rho}+\frac{p\ddot{E}_o(0,\rho)}{2\rho}
\rightarrow \frac{\dot{E}_o(0,\rho)}{\rho} \quad\textsl{uniformly
for ${\rho\in[0,1]}.$}\label{eq: o0}
\end{equation}
From the first-order calculation, we know that
$\dot{E}_o(0,\rho)=\frac{\rho}{1+\rho}.$

We prove $\lim_{p\rightarrow 0}\rho(p)=\rho^*$ using a formal
definition of the limit. For any $\epsilon_0>0,$ we show that we
can find $\delta>0$ such that for all $p<\delta,$ we always have
$$
|\rho(p)-\rho^*|<\epsilon_0.
$$

To see this, define
$$
\epsilon=(1-\sqrt{z})^2-\min(g(\rho^*-\epsilon_0),g(\rho^*+\epsilon_0)),
$$
where
$$g(\rho)=-\frac{z}{\rho}+\frac{1}{1+\rho}.$$

Now we use (\ref{eq: o0}) here. For this $\epsilon$, we can find
$\delta'>0$ such that for any $\rho\in [0,1]$ and for all
$p<\delta',$ such that
$$
\left|\frac{\dot{E}_o(0,\rho)}{\rho}+
\frac{p\ddot{E}_o(0,\rho)}{2\rho}- \frac{1}{1+\rho}\right|\le
\frac{\epsilon}{2}.
$$
Thus, we have
$$
\sup_{0\le\rho\le 1}
-\frac{z}{\rho}+\frac{\dot{E}_o(0,\rho)}{\rho}
+\frac{p\ddot{E}_o(0,\rho)}{2\rho}
>\sup_{0\le\rho\le 1}
-\frac{z}{\rho}+\frac{1}{1+\rho}-\frac{\epsilon}{2}
=(1-\sqrt{z})^2-\frac{\epsilon}{2}.
$$
On the other hand, we also have
\begin{eqnarray*}
&&\sup_{0\le\rho\le 1}
-\frac{z}{\rho}+\frac{\dot{E}_o(0,\rho)}{\rho}
+\frac{p\ddot{E}_o(0,\rho)}{2\rho}
\\
&=& -\frac{z}{\rho(p)}+\frac{1}{1+\rho(p)}
+\frac{p\ddot{E}_o(0,\rho(p))}{2\rho(p)} \\
&\le& g(\rho(p))+\frac{pM}{2},
\end{eqnarray*}
where $M$ is the upper bound for $\frac{\ddot{E}_o(0,\rho)}{\rho}$
for $\rho\in[0,1].$ We choose
$\delta=\min(\delta',\frac{\epsilon}{M}),$ then for all
$p<\delta,$ we have $\frac{pM}{2}\le \frac{\epsilon}{2}.$ Further,
$$
g(\rho(p))\ge (1-\sqrt{z})^2-\frac{\epsilon}{2}-\frac{pM}{2}\ge
(1-\sqrt{z})^2-\epsilon.
$$

From the definition of $\epsilon,$ we must have
$$
|\rho(p)-\rho^*|<\epsilon_0,
$$
which finishes the proof of this part.

\section{Proof of Lemma~\ref{lem: temp18 0}}\label{sec: temp18 0}

The first-order calculation gives us
$$
{\dot{E}}_o(0,\rho)=\frac{\rho}{1+\rho}.
$$ Thus,
\begin{eqnarray}
&&\frac{\frac{E_o(p,\rho)}{p\rho}-\frac{\dot{E}_o(0,\rho)}{\rho}}{p}\nonumber
\\ &=&\frac{-\ln\left(\inf_{\{q_p\}\in \tilde{\mathcal G}(p)}
\inf_{\beta\ge 0}\int
\alpha(y)^{1+\rho}dy\right)-\frac{\rho p}{1+\rho}}{\rho p^2}\nonumber \\
&=&\frac{-\ln\left(e^{\frac{\rho p}{1+\rho}} \inf_{\{q_p\}\in
\tilde{\mathcal G}(p)}
\inf_{\beta\ge 0}\int\alpha(y)^{1+\rho} dy\right)}{\rho p^2} \nonumber \\
&=& \frac{\ln\frac{1}{e^{\frac{\rho p}{1+\rho}} \inf_{\{q_p\}\in
\tilde{\mathcal G}(p)}
\inf_{\beta \ge 0}\int\alpha(y)^{1+\rho}dy}}{\rho p^2} \nonumber \\
&\le& \frac{-1+\frac{1}{e^{\frac{\rho p}{1+\rho}} \inf_{\{q_p\}\in
\tilde{\mathcal G}(p)}
\inf_{\beta\ge 0}\int\alpha(y)^{1+\rho}dy}}{\rho p^2} \nonumber \\
&=& \frac{-e^{\frac{\rho p}{1+\rho}}\inf_{\{q_p\}\in
\tilde{\mathcal G}(p)} \inf_{\beta\ge 0} \int \alpha(y)^{1+\rho}dy
+1}{\rho p^2}\frac{1}{e^{\frac{\rho p}{1+\rho}}\inf_{\{q_p\}\in
\tilde{\mathcal G}(p)}\inf_{\beta \ge 0} \int\alpha(y)^{1+\rho}dy}
\nonumber \\
&\le &\frac{-e^{\frac{\rho p}{1+\rho}}\inf_{\{q_p\}\in
\tilde{\mathcal G}(p)}\inf_{\beta\ge 0}\int \alpha(y)^{1+\rho}dy
+1}{\rho p^2 e^{\frac{\rho p}{1+\rho}}}\label{eq: l00} \\
&=& \frac{-\inf_{\{q_p\}\in \tilde{\mathcal G}(p)}\inf_{\beta\ge
0}\int \alpha(y)^{1+\rho}dy +e^{-\frac{\rho p}{1+\rho}}}{\rho p^2
}.\nonumber
\end{eqnarray}
The inequality (\ref{eq: l00}) is true because Lemma~\ref{lem:
upper bound} implies
$$
\inf_{\{q_p\}\in \tilde{\mathcal G}(p)}\inf_{\beta\ge 0}\int
\alpha(y)^{1+\rho}dy=e^{-E_o(p,\rho)}\ge  e^{-\frac{\rho
p}{1+\rho}},
$$
which leads to
$$
-e^{\frac{\rho p}{1+\rho}} \inf_{\{q_p\}\in \tilde{\mathcal
G}(p)}\inf_{\beta\ge 0} \int \alpha(y)^{1+\rho} dy +1\le 0.$$ On
the other hand,
\begin{eqnarray*}
\inf_{\{q_p\}\in \tilde{\mathcal G}(p)}\inf_{\beta\ge 0} \int
\alpha(y)^{1+\rho} dy&\le& \inf_{\{q_p\}\in \tilde{\mathcal G}(p)}
\int \alpha(y)^{1+\rho} dy|_{\beta=0}\\
&=& \inf_{\{q_p\}\in \tilde{\mathcal G}(p)} \int \left(\sum_{k}
q_k f(y|x_k)^\frac{1}{1+\rho}\right)^{1+\rho} dy\\
&\le& \inf_{\{q_p\}\in \tilde{\mathcal G}(p)} \int \left(\sum_{k}
q_k f(y|x_k)\right) dy \\
&=& 1.
\end{eqnarray*}
These two bounds together give us (\ref{eq: l00}).

\section{Proof of Lemma~\ref{lem: T2 0}}\label{sec: T2 0}

First we check that
\begin{equation}
\int f_w(y) M^2(y) dy
=E\left[e^{\beta^*(|x_1|^2+|x_2|^2-2p)}e^{-\theta(|x_1|^2+|x_2|^2)}
e^{\frac{2Re(x_1x_2^*)}{(1+\rho^*)^2}}\right],
\end{equation}
and thus
\begin{eqnarray*}
\int f_w(y) T^2(y) dy&=& \int f_w(y) (M(y)-1)^2 dy\\
&=&
E\left[e^{\beta^*(|x_1|^2+|x_2|^2-2p)}e^{-\theta(|x_1|^2+|x_2|^2)}
\left(e^{\frac{2Re(x_1x_2^*)}{(1+\rho^*)^2}}-1\right)\right]
+\left( E\left[e^{\beta^* (|x|^2-p)} e^{-\theta
|x|^2}\right]-1\right)^2.
\end{eqnarray*}
Since ${\tilde E}_o(p,q_p,\rho)\ge {\tilde E}_o(p,QPSK,\rho),$ and
$$
{\tilde E}_o(p,q_p,\rho)=-\ln \int \alpha(y)^{1+\rho} dy \le
-(1+\rho)\ln E\left[e^{\beta^* (|x|^2-p)} e^{-\theta
|x|^2}\right],
$$
we have
$$
E\left[e^{\beta^* (|x|^2-p)} e^{-\theta |x|^2}\right] \le
e^{-\frac{{\tilde E}_o(p,QPSK,\rho)}{1+\rho}}.
$$
As we will show later, $\frac{{\tilde E}_o(p,QPSk,\rho)}{\rho p}$ converges
to $\frac{1}{1+\rho}$ uniformly. In other words, we can write
${\tilde E}_o(p,QPSK,\rho)$ as $\frac{\rho p}{1+\rho}+\rho\delta(p),$ where
$\frac{\delta(p)}{p}$ goes to zero uniformly for all $\rho$ as $p$
goes to $0.$ Thus,
$$
E\left[e^{\beta^* (|x|^2-p)} e^{-\theta |x|^2}\right] \le
e^{-\frac{\rho p}{(1+\rho)^2}+\frac{\rho}{1+\rho} \delta(p)}.
$$

Note we should always have
$$
E[\left[e^{\beta^* (|x|^2-p)} e^{-\theta |x|^2}\right] \le 1,
$$
for the optimizing $\beta^*.$ This can be seen by the following
sequence of inequalities:
\begin{eqnarray*}
&&(E\left[e^{\beta^* (|x|^2-p)} e^{-\theta
|x|^2}\right])^{1+\rho} \\
&\le& \inf_{\beta \ge 0} \int \alpha^{1+\rho} dy \\
&\le& \left. \int \alpha^{1+\rho} dy\right|_{\beta=0} \\
&=& \int \left( \sum_k q_k f(y|x_k)^{\frac{1}{1+\rho}}
\right)^{1+\rho} dy \\
&\le& \int \sum_k q_k f(y|x_k) dy \\
&=& 1.
\end{eqnarray*}
Thus,
\begin{eqnarray*}
&&\left( E\left[e^{\beta^* (|x|^2-p)} e^{-\theta
|x|^2}\right]-1\right)^2 \\
&\ge& \left( e^{-\frac{\rho p}{(1+\rho)^2}+\frac{\rho}{1+\rho}
\delta(p)}-1\right)^2 \\
&\ge& \left\{\frac{\rho p}{(1+\rho)^2}-\frac{\rho}{1+\rho}
\delta(p) -\frac{\left(\frac{\rho
p}{(1+\rho)^2}-\frac{\rho}{1+\rho}
\delta(p)\right)^2}{2}\right\}^2 \\
&=&\theta^2 p^2+\delta(p^2).
\end{eqnarray*}

On the other hand, we have
\begin{eqnarray*}
&&E\left[e^{\beta^*(|x_1|^2+|x_2|^2-2p)}e^{-\theta(|x_1|^2+|x_2|^2)}
\left(e^{\frac{2Re(x_1x_2^*)}{(1+\rho)^2}}-1\right)\right]\\
&\ge&
E\left[e^{\beta^*(|x_1|^2+|x_2|^2-2p)}e^{-\theta(|x_1|^2+|x_2|^2)}
\frac{2Re(x_1x_2^*)^2}{(1+\rho)^4}\right]
\\
&=&E\left[e^{\beta^*(|x_1|^2+|x_2|^2-2p)}e^{-\theta(|x_1|^2+|x_2|^2)}
\frac{2(x_{1r}^2x_{2r}^2+x_{1c}^2x_{2c}^2+2x_{1r}x_{1c}x_{2r}x_{2c})}{(1+\rho)^4}
\right]\\
&\ge&\frac{2
}{(1+\rho)^4}\left\{(E[e^{\beta^*(|x_1|^2-p)}e^{-\theta
|x_1|^2}x_{1r}^2])^2 +(E[e^{\beta^*(|x_1|^2-p)}e^{-\theta
|x_1|^2}x_{1c}^2])^2\right\}\\
&\ge& \frac{\left(E[e^{\beta^*(|x_1|^2-p)}e^{-\theta
|x_1|^2}(x_{1r}^2+x_{1c}^2)]\right)^2}{(1+\rho)^4}\\
&=& \frac{\left(E[e^{\beta^*(|x_1|^2-p)}e^{-\theta
|x_1|^2}|x_1|^2]\right)^2}{(1+\rho)^4} \\
&\ge& \frac{\left(E[(1+\beta^*(|x_1|^2-p) -\theta
|x_1|^2)|x_1|^2]\right)^2}{(1+\rho)^4} \\
&\ge& \frac{\left(p -\theta E[|x_1|^4]\right)^2}{(1+\rho)^4} \\
&\ge& \frac{\left(p -\theta K_m^2 p^{1+2\alpha} \right)^2}{(1+\rho)^4} \\
&\ge& \frac{p^2 -2\theta K_m^2 p^{2+2\alpha}}{(1+\rho)^4}\\
&=&\frac{p^2}{(1+\rho)^4}+\delta(p^2).
\end{eqnarray*}
In the above equations, $x_{ir}$, $x_{ic}$ denote the real part
and imaginary part of the random variable $x_i$, $i=1,2$.
\section{Proof of Lemma~\ref{lem: T3 0}}\label{sec: appendix 3}
To prove Lemma~\ref{lem: T3 0}, we first establish two other
lemmas. The first lemma shows that we can restrict ourselves to
considering distributions which are symmetric around $0.$ Define
$\Gamma(q)=\int \alpha(y)^{1+\rho} dy$.

\begin{lemma}
Given any distribution $q(x)\in {\tilde\mathcal F}(p),$ we can
find a symmetric distribution $q_e(x)\in {\tilde\mathcal F}(p),$
i.e., $q_e(x)=q_e(-x)\enskip \forall x,$ such that $\Gamma(q_e)
\leq \Gamma(q).$
\end{lemma}
Proof: We first compute $\Gamma(\cdot)$ for $q(-x)$ and show that
it is the same as $\Gamma(q).$
\begin{eqnarray*}
&&\int\left(\int
q(-x)e^{\beta(|x|^2-p)}f_w(y-x)^{\frac{1}{1+\rho}}dx
\right)^{1+\rho} dy\nonumber\\
&=&\int\left(\int
q(x)e^{\beta(|-x|^2-p)}f_w(y+x)^{\frac{1}{1+\rho}}dx
\right)^{1+\rho} dy\\
&=&\int\left(\int
q(x)e^{\beta(|x|^2-p)}f_w(-y+x)^{\frac{1}{1+\rho}}dx
\right)^{1+\rho} dy\\
&=&\int\left(\int
q(x)e^{\beta(|x|^2-p)}f_w(y-x)^{\frac{1}{1+\rho}}dx
\right)^{1+\rho} dy \\
&=&\Gamma(q).
\end{eqnarray*}
For $\rho\in[0,1]$, it is easy to see that $\int
\alpha(y)^{1+\rho} dy$ is a convex function of $q(x)$ for a fixed
$\beta$ . Thus if we choose $q_e(x)=\frac{1}{2}(q(x)+q(-x))$, the
power constraint will be still valid and we have
$$
\Gamma(q_e(x))=\int \left(\int q_e(x)
e^{\beta(|x|^2-p)}f_w(y-x)^{\frac{1}{1+\rho}} dx\right)^{1+\rho}
dy \le \frac{1}{2}(\Gamma(q)+\Gamma(q))=\Gamma(q).
$$
\hfill $\diamond$

The second lemma provides an upper bound for
$E[e^{\beta^*(|x|^2-p)}e^{-\theta |x|^2}|x|^2],$ which is a key
term in the proof of Lemma~\ref{lem: T3 0}.

\begin{lemma}
For any input distribution $\{q_p\}$ which has mean variance $p,$
let $\beta^*$ be the optimizing $\beta$ as in (\ref{eq: int
alpha}). We must have
\begin{equation}
E[e^{\beta^*(|x|^2-p)}e^{-\theta|x|^2}|x|^2]\le pe^{\theta
p}.\label{eq: asf}
\end{equation}
\end{lemma}
Proof: Denote
$$
h(\beta)=\int \left(\sum q_k e^{\beta(|x_k|^2-p)}
f(y|x_k)^\frac{1}{1+\rho}\right)^{1+\rho} dy.
$$
If $\beta^*$ is the optimizing $\beta,$ applying the Kuch-Tucker
condition here, we must have
$$
\beta^* h'(\beta^*)=0,
$$
which yields $\beta^*=0$ or
$$
\int \left(\sum_k q_k e^{\beta^*(|x_k|^2-p)}
f(y|x_k)^\frac{1}{1+\rho}\right)^{\rho} \sum_k q_k
e^{\beta^*(|x_k|^2-p)} f(y|x_k)^\frac{1}{1+\rho} (|x_k|^2-p)dy=0,
$$
which can be simplified as
\begin{equation}
p\int \alpha(y)^{1+\rho} dy=\int \alpha(y)^\rho \gamma(y) dy.
\label{eq: dfg}
\end{equation}
Here we let
\begin{eqnarray*}
\alpha(y)&=&\sum_k q_k e^{\beta^*(|x_k|^2-p)}
f(y|x_k)^\frac{1}{1+\rho};\\
\gamma(y)&=&\sum_k q_k e^{\beta^*(|x_k|^2-p)}
f(y|x_k)^\frac{1}{1+\rho} |x_k|^2.
\end{eqnarray*}

If $\beta^*=0,$ (\ref{eq: asf}) is trivial.

If $\beta^*>0,$ we derive (\ref{eq: asf}) using (\ref{eq: dfg}).
Note that
\begin{eqnarray*}
\int \alpha(y)^\rho \gamma(y) dy &\ge& \int \sum_k q_k
e^{\beta^*\rho (|x_k|^2-p)} f(y|x_k)^{\frac{\rho}{1+\rho}}
\gamma(y) dy\\
&=&\sum_k q_k e^{\beta^*\rho(|x_k|^2-p)} \sum_{l} q_l
e^{\beta^*(|x_l|^2-p)} |x_l|^2 \int f(y|x_k)^{\frac{\rho}{1+\rho}}
f(y|x_l)^{\frac{1}{1+\rho}} dy\\
&=& \sum_l q_l e^{\beta^*(|x_l|^2-p)} |x_l|^2 \sum_k q_k
e^{\beta^*\rho(|x_k|^2-p)} e^{-\theta |x_k-x_l|^2}\\
&\ge& \sum_l q_l e^{\beta^*(|x_l|^2-p)} |x_l|^2 e^{\sum_k q_k
\{\beta^*\rho(|x_k|^2-p)-\theta |x_k-x_l|^2\}}\\
&\ge& \sum_l q_l e^{\beta^*(|x_l|^2-p)} |x_l|^2 e^{-\theta
p}e^{-\theta |x_l|^2} \\
&=& e^{-\theta p} E[e^{\beta^*(|x|^2-p)}e^{-\theta|x|^2}|x|^2].
\end{eqnarray*}

On the other hand, as we have shown before,
\begin{eqnarray*}
\int \alpha(y)^{1+\rho} dy &=& \inf_{\beta\ge 0} \int \left(\sum
q_k e^{\beta(|x_k|^2-p)} f(y|x_k)^\frac{1}{1+\rho}\right)^{1+\rho}
dy\le 1.
\end{eqnarray*}

Thus, we must have
\begin{equation}
E[e^{\beta^*(|x|^2-p)}e^{-\theta|x|^2}|x|^2]\le e^{\theta p} \int
\alpha(y)^\rho \gamma(y) dy = p e^{\theta p} \int
\alpha(y)^{1+\rho} dy \le p e^{\theta p}.
\end{equation}
\hfill $\diamond$

Now we prove Lemma~\ref{lem: T3 0}.
\begin{eqnarray*}
\int f_w(y)T^3(y)dy&=&\int f_w(y)( M(y)-1)^3dy\\
        &=&\int
f_w(y)( M^3(y)-3 M^2(y)+3 M(y)-1)dy.
\end{eqnarray*}
It is easy to check that
\begin{eqnarray*}
\int f_w(y) M(y)dy&=&E\left[e^{\beta^*(|x_1|^2-p)}e^{-\theta
|x_1|^2}\right];\\
\int f_w(y)
M^2(y)dy&=&E\left[e^{\beta^*(|x_1|^2+|x_2|^2-2p)}e^{-\theta(|x_1|^2+|x_2|^2)}e^
{\frac{2Re(x_1x_2^*)}{(1+\rho)^2}}\right];\\
\int f_w(y) M^3(y)dy&=&
E\left[e^{\beta^*(|x_1|^2+|x_2|^2+|x_3|^2-3p)}e^{-\theta(|x_1|^2+|x_2|^2+|x_3|^2)}
e^{\frac{2Re(x_1x_2^*+x_1x_3^*+x_2x_3^*)}{(1+\rho)^2}}\right],\nonumber\\
\end{eqnarray*}
where $x_1$, $x_2$ and $x_3$ are i.i.d. random variables with
distribution $\{q_p(x)\}.$ Thus, after some manipulations, we have
\begin{eqnarray}
&&\int
f_w(y)T^3(y)dy=\left(E\left[e^{\beta^*(|x_1|^2-p)}e^{-\theta
|x_1|^2}\right]-1\right)^3 \nonumber\\
&&-3E\left[e^{\beta^*(|x_1|^2+|x_2|^2-2p)}e^{-\theta(|x_1|^2+|x_2|^2)}\left(e^{
\frac{2Re(x_1x_2^*)}{(1+\rho)^2}}-1\right)\right]\nonumber\\
&&+E\left[e^{\beta^*(|x_1|^2+|x_2|^2+|x_3|^2-3p)}e^{-\theta(|x_1|^2+|x_2|^2+|x_3|^2)}
\left(e^{
\frac{2Re(x_1x_2^*+x_1x_3^*+x_2x_3^*)}{(1+\rho)^2}}-1\right)\right].
\label{eq: temp15 0}
\end{eqnarray}

From the proof in Lemma~\ref{lem: T2 0}, we know
$$
e^{-\theta p}-1\le E\left[e^{\beta^*(|x_1|^2-p)}e^{-\theta
|x_1|^2}\right]-1\le 0,
$$
and thus, we must have
$$
\left|E\left[\exp\left\{(\beta^*-\theta)(|x_1|^2-p)\right\}\right]e^{-\theta
p}-1\right|^3\le (1-e^{-\theta p})^3\le \theta^3 p^3.
$$

On the other hand, we expand the second and third term in the RHS
of (\ref{eq: temp15 0}) as follows:
\begin{eqnarray}
&&E\left[e^{\beta^*(|x_1|^2+|x_2|^2-2p)}e^{-\theta(|x_1|^2+|x_2|^2)}
\left(e^
{\frac{2Re(x_1x_2^*)}{(1+\rho)^2}}-1\right)\right]\nonumber\\
=&&\sum_{k=1}^\infty
E\left[e^{\beta^*(|x_1|^2+|x_2|^2-2p)}e^{-\theta(|x_1|^2+|x_2|^2)}
\frac{(2Re(x_1x_2^*))^k}{(1+\rho)^{2k}k!}\right],\label{eq:
temp12}
\end{eqnarray}
and
\begin{eqnarray}
&&E\left[e^{\beta^*(|x_1|^2+|x_2|^2+|x_3|^2-3p)}e^{-\theta(|x_1|^2+|x_2|^2+|x_3|^2)}
\left(e^{
\frac{2Re(x_1x_2^*+x_1x_3^*+x_2x_3^*)}{(1+\rho)^2}}-1\right)\right]
\nonumber\\
=&&\sum_{k=1}^\infty
E\left[e^{\beta^*(|x_1|^2+|x_2|^2+|x_3|^2-3p)}e^{-\theta(|x_1|^2+|x_2|^2+|x_3|^2)}
\frac{2^{k}(Re(x_1x_2^*)+Re(x_1x_3^*)+Re(x_2x_3^*))^k}{(1+\rho)^{2k}k!}\right]\nonumber
\\
 =&&\sum_{k=1}^\infty \sum_{l+m+n=k}
E\left[e^{\beta^*(|x_1|^2+|x_2|^2+|x_3|^2-3p)}e^{-\theta(|x_1|^2+|x_2|^2+|x_3|^2)}
{ \frac{2^{k}C_{lmn}^{(k)} Re(x_1x_2^*)^l Re(x_1x_3^*)^m
Re(x_2x_3^*)^n}{(1+\rho)^{2k}k!}}\right],\nonumber
\end{eqnarray}
where $C_{lmn}^{(k)}$ is a non-negative constant independent of
$p.$

It is straightforward to check to following, using the above two
expansions:
\begin{eqnarray*}
&&E\left[e^{\beta^*(|x_1|^2+|x_2|^2+|x_3|^2-3p)}e^{-\theta(|x_1|^2+|x_2|^2+|x_3|^2)}
\left(e^{
\frac{2Re(x_1x_2^*+x_1x_3^*+x_2x_3^*)}{(1+\rho)^2}}-1\right)\right]\\
&&-3E\left[e^{\beta^*(|x_1|^2+|x_2|^2-2p)}e^{-\theta(|x_1|^2+|x_2|^2)}\left(e^{
\frac{2Re(x_1x_2^*)}{(1+\rho)^2}}-1\right)\right]\\
&=& \sum_{k=1}^\infty \sum_{
\begin{array}{l}
l+m+n=k;\\
l,m,n<k
\end{array}}
E\left[e^{\beta^*(|x_1|^2+|x_2|^2+|x_3|^2-3p)}e^{-\theta(|x_1|^2+|x_2|^2+|x_3|^2)}
{ \frac{2^{k}C_{lmn}^{(k)} Re(x_1x_2^*)^l Re(x_1x_3^*)^m
Re(x_2x_3^*)^n}{(1+\rho)^{2k}k!}}\right]\\
&&+3\sum_{k=1}^\infty
E\left[e^{\beta^*(|x_1|^2+|x_2|^2-2p)}e^{-\theta(|x_1|^2+|x_2|^2)}
\frac{(2Re(x_1x_2^*))^k}{(1+\rho)^{2k}k!}\right]
\left\{E[e^{\beta^*(|x_3|^2-p)}e^{-\theta |x_3|^2}]-1\right\}.
\end{eqnarray*}

Next, we bound the two terms above separately, using the bound
that $Re(z)\le |z|.$ Note that for symmetric distributions, it is
easy to see that all the $k$ odd terms will vanish. Thus, we can
remove the term with $k=1.$
\begin{eqnarray*}
&&\left|\sum_{k=1}^\infty
E\left[e^{\beta^*(|x_1|^2+|x_2|^2-2p)}e^{-\theta(|x_1|^2+|x_2|^2)}
\frac{(2Re(x_1x_2^*))^k}{(1+\rho)^{2k}k!}\right]
\left\{E[e^{\beta^*(|x_3|^2-p)}e^{-\theta
|x_3|^2}]-1\right\}\right|\\
&\le& \sum_{k=2}^\infty\left|
E\left[e^{\beta^*(|x_1|^2+|x_2|^2-2p)}e^{-\theta(|x_1|^2+|x_2|^2)}
\frac{(2Re(x_1x_2^*))^k}{(1+\rho)^{2k}k!}\right]\right|
\left|\left\{E[e^{\beta^*(|x_3|^2-p)}e^{-\theta
|x_3|^2}]-1\right\}\right|\\
&\le& \sum_{k=2}^\infty
E\left[e^{\beta^*(|x_1|^2+|x_2|^2-2p)}e^{-\theta(|x_1|^2+|x_2|^2)}
\frac{2^k|x_1|^k|x_2|^k}{(1+\rho)^{2k}k!}\right] (1-e^{-\theta p})
\\
&\le& \theta p \sum_{k=2}^\infty \frac{2^k}{(1+\rho)^{2k}k!}\left(
E\left[e^{\beta^*(|x_1|^2-p)}e^{-\theta |x_1|^2} |x_1|^k
\right]\right)^2\\
&\le& \theta p \sum_{k=2}^\infty \frac{2^k
(K_m p^\alpha)^{2(k-2)}}{(1+\rho)^{2k}k!}\left(
E\left[e^{\beta^*(|x_1|^2-p)}e^{-\theta |x_1|^2} |x_1|^2
\right]\right)^2 \\
&\le& 4\theta p e^{2K_m^2}\left(
E\left[e^{\beta^*(|x_1|^2-p)}e^{-\theta |x_1|^2} |x_1|^2
\right]\right)^2 \\
&\le& 4\theta e^{2\theta p} e^{2K_m^2} p^3.
\end{eqnarray*}

Similarly, for the other term, we can also remove the term where
$k$ is odd. Actually, we can do more. For example, when $k=2,$
since at least two of $l,m,n$ are required to be non-zero, we must
have two of them are $1,$ while the other is $0.$ It can be easily
seen the contribution of this term is also zero, for symmetric
distributions. Thus, we remove the terms for both $k=1$ and $k=2.$
\begin{eqnarray*}
&&\sum_{k=1}^\infty \sum_{
\begin{array}{l}
l+m+n=k;\\
l,m,n<k
\end{array}}
E\left[e^{\beta^*(|x_1|^2+|x_2|^2+|x_3|^2-3p)}e^{-\theta(|x_1|^2+|x_2|^2+|x_3|^2)}
{ \frac{2^{k}C_{lmn}^{(k)} Re(x_1x_2^*)^l Re(x_1x_3^*)^m
Re(x_2x_3^*)^n}{(1+\rho)^{2k}k!}}\right]\\
&\le& \sum_{k=3}^\infty \frac{2^{k}}{(1+\rho)^{2k}k!}\sum_{
\begin{array}{l}
l+m+n=k;\\
l,m,n<k
\end{array}}
C_{lmn}^{(k)}E\left[e^{\beta^*(|x_1|^2+|x_2|^2+|x_3|^2-3p)}e^{-\theta(|x_1|^2+|x_2|^2+|x_3|^2)}
|x_1|^{l+m}|x_2|^{m+n}|x_3|^{m+n}\right]\\
&=&\sum_{k=3}^\infty \frac{2^{k}}{(1+\rho)^{2k}k!}\sum_{
\begin{array}{l}
l+m+n=k;\\
l,m,n<k
\end{array}}
C_{lmn}^{(k)} E\left[e^{\beta^*(|x_1|^2-p)}e^{-\theta |x_1|^2}
|x_1|^{l+m}\right]*\nonumber\\
&&E\left[e^{\beta^*(|x_2^2-p)}e^{-\theta |x_2|^2}
|x_2|^{l+n}\right]* E\left[e^{\beta^*(|x_3|^2-p)}e^{-\theta
|x_3|^2} |x_3|^{m+n}\right]\\
&\le&\sum_{k=3}^\infty
\frac{2^{k}K_m^{2k-6}}{(1+\rho)^{2k}k!}\sum_{
\begin{array}{l}
l+m+n=k;\\
l,m,n<k
\end{array}}
C_{lmn}^{(k)} \left(E\left[e^{\beta^*(|x_1|^2-p)}e^{-\theta
|x_1|^2} |x_1|^2\right]\right)^3\\
&\le& \sum_{k=3}^\infty \frac{6^{k}K_m^{2k-6}}{(1+\rho)^{2k}k!}
\left(E\left[e^{\beta^*(|x_1|^2-p)}e^{-\theta |x_1|^2}
|x_1|^2\right]\right)^3\\
&\le& 216 e^{6K_m^2} \left(E\left[e^{\beta^*(|x_1|^2-p)}e^{-\theta
|x_1|^2} |x_1|^2\right]\right)^3 \\
&\le& 216 e^{6K_m^2} e^{3\theta p} p^3.
\end{eqnarray*}

Combining all these bounds, we have
\begin{eqnarray*}
\left| \int f_w(y) T^3(y) dy \right| &\le& \theta^3 p^3 + 12\theta
e^{2\theta p} e^{2K_m^2} p^3+ 216 e^{6K_m^2} e^{3\theta p} p^3\\
&\le& Ce^{3\theta p} p^3,
\end{eqnarray*}
where $C$ is a constant, which is independent of $\rho$ and
independent of the choice of input distributions, as far as it is
in ${\tilde\mathcal F}(p).$

\section{Proof of Lemma~\ref{lem: mean zero optimal}}\label{sec:
proof all}

To show this, we need to check (\ref{eq: first order condition})
for a sequence of mean-zero input distribution $q_p\in
{\tilde\mathcal F}(p).$ Since it is always true that
$$
\limsup_{p\rightarrow 0} \frac{{\tilde E}_o(p,q_p,\rho^*)}{p}\le
\frac{\rho^*}{1+\rho^*},
$$
it suffices to show that
$$
\liminf_{p\rightarrow 0} \frac{{\tilde E}_o(p,q_p,\rho^*)}{p}\ge
\frac{\rho^*}{1+\rho^*}.
$$
Note
\begin{eqnarray*}
{\tilde E}_o(p,q_p,\rho^*)&=&\sup_{\beta\ge 0} -\ln \int
\alpha(y)^{1+\rho^*}
dy\\
&=& \sup_{\beta\ge 0} -\ln \int f_w(y) (1+T(y))^{1+\rho^*} dy.
\end{eqnarray*}
To achieve a lower bound, we choose
$\beta=\theta=\frac{\rho^*}{1+\rho^*}.$ Further, we use the following
inequality
$$
(1+t)^{1+\rho^*}\le 1+(1+\rho^*)t+\frac{\rho^*(1+\rho^*)}{2} t^2.
$$
This leads to
$$
{\tilde E}_o(p,q_p,\rho^*) \ge -\ln \int f_w(y) (1+(1+\rho^*)T(y)
+ \frac{\rho^*(1+\rho^*)}{2} T^2(y)) dy.
$$

When $\beta=\theta,$ it can be shown that
$$
\int f_w(y) (1+(1+\rho^*)T(y)) dy=-\rho^*+(1+\rho^*)e^{-\theta p},
$$
and
$$
\int f_w(y) T^2(y) dy=1-2e^{-\theta p} + E\left[e^{\frac{2Re(x_1
x_2^*)}{(1+\rho^*)^2}} \right]e^{-2\theta p},
$$
where $x_1$ and $x_2$ are i.i.d random variables distributed
according to $q_p(x).$

Next we claim
$$
\lim_{p\rightarrow 0} \frac{\int f_w(y) T^2(y)dy}{p}=0.
$$

Since $\lim_{p\rightarrow 0} \frac{(1-e^{-\theta p})^2}{p}=0,$ it
suffices to show
$$
\lim_{p\rightarrow 0} \frac{E\left[e^{\frac{2Re(x_1
x_2^*)}{(1+\rho^*)^2}} \right]-1}{p}=0.
$$
Using the assumption that $q_p(x)$ is symmetric around $0$ and
$$
|x|_{max}<K_m p^{\alpha},
$$
we can show this following a similar procedure as in the proof of
Lemma~\ref{lem: T3 0}.

Thus, we have
\begin{eqnarray*}
\liminf_{p\rightarrow 0} \frac{{\tilde E}_o(p,q_p,\rho^*)}{p}&\ge&
\liminf_{p\rightarrow 0} \frac{-\ln(-\rho^*+(1+\rho^*)e^{-\theta
p}+o(p))}{p} \\
&=& \liminf_{p\rightarrow 0}
\frac{-\ln(1-\frac{\rho^* p}{1+\rho^*}+o(p))}{p}\\
&=&\frac{\rho^*}{1+\rho^*}.
\end{eqnarray*}

\section{BPSK and QPSK for AWGN channels} \label{sec:
BPSK and QPSK}

Since for both BPSK and QPSK, we have $|x|^2=p$ with probability
$1$, the power constraint parameter $\beta$ does not play a role
here and ${\tilde E}_o(p,q_p,\rho)$ can be simplified to
$$
{\tilde E}_o(p,q_p,\rho)=-\ln\int\alpha(y)^{1+\rho} dy,
$$
with
$$
\alpha(y)=\int q_p(x)f_w(y|x)^{\frac{1}{1+\rho}}dx.
$$

Again, we use the two inequalities which have been very helpful to us
in the general first and second order calculations:
\begin{eqnarray}
(1+t)^{1+\rho}&\le& 1+(1+\rho)t+\frac{\rho(1+\rho)}{2} t^2;\label{eq: t2}\\
(1+t)^{1+\rho}&\ge& 1+(1+\rho)t+\frac{\rho(1+\rho)}{2} t^2 -\frac{\rho(1+\rho)(1-\rho)}{6} t^3.
\label{eq: t3}
\end{eqnarray}

We write $\int \alpha(y) dy$ as follows
\begin{equation}
\int \alpha(y) dy= \int f_w(y) (1+T(y))^{1+\rho} dy,
\end{equation}
where $T(y)$ denotes
$$
T(y)=\sum_{k} q_k \left(\frac{f(y|x)}{f(y|0)}\right)^{\frac{1}{1+\rho}}-1.
$$

It is easy to check for BPSK or QPSK, we have
\begin{eqnarray*}
\int f_w(y) T(y) dy&=& e^{-\theta p}-1;\\
\int f_w(y) T^2(y) dy &=& (e^{-\theta p}-1)^2+e^{-2\theta p}
E[e^{\frac{2Re(x_1x_2^*)}{(1+\rho)^2}}-1];\\
\int f_w(y) T^3(y) dy &=& (e^{-\theta p}-1)^3 + e^{-3\theta p}
E[e^{\frac{2Re(x_1x_2^*)+2Re(x_1x_3^*)+2Re(x_2x_3^*)}{(1+\rho)^2}}-1]
-3 e^{-2\theta p} E[e^{\frac{2Re(x_1x_2^*)}{(1+\rho)^2}}-1].
\end{eqnarray*}

Further, for BPSK, we can calculate that
\begin{equation}
E[e^{\frac{2Re(x_1x_2^*)}{(1+\rho)^2}}]= \frac{1}{2} \left(e^{\frac{2p}{(1+\rho)^2}}
+e^{-\frac{2p}{(1+\rho)^2}}-2\right)=1+\frac{2p^2}{(1+\rho)^4}+\delta(p^2).
\end{equation}
and
\begin{eqnarray*}
E[e^{\frac{2Re(x_1x_2^*)}{(1+\rho)^2}}]&=& 1+\frac{2p^2}{(1+\rho)^4}+\delta(p^2);\\
E[e^{\frac{2Re(x_1x_2^*)+2Re(x_1x_3^*)+2Re(x_2x_3^*)}{(1+\rho)^2}}]&=&1+\frac{6p^2}{(1+\rho)^4}+\delta(p^2),
\end{eqnarray*}
which further yield an upper bound and lower bound for $\int \alpha(y) dy,$
\begin{eqnarray*}
\int \alpha(y) dy &\le& 1+(1+\rho) \int f_w(y) T(y) dy +\frac{\rho(1+\rho)}{2} \int f_w(y) T^2(y) dy\\
&=& 1+(1+\rho)(e^{-\theta p}-1) + \frac{\rho(1+\rho)}{2}\left\{ (e^{-\theta p} -1)^2 +
\frac{2p^2}{(1+\rho)^4}+\delta(p^2)\right\}\\
&\le& 1+(1+\rho)(-\theta p+\frac{\theta^2 p^2}{2}) + \frac{\rho(1+\rho)}{2}\left\{ \theta^2p^2 +
\frac{2p^2}{(1+\rho)^4}+\delta(p^2)\right\}\\
&=& 1-\frac{\rho}{1+\rho}p+\frac{\rho^3+\rho^2+2\rho}{(1+\rho)^3} \frac{p^2}{2} +\rho \delta(p^2);\\
\int \alpha(y) dy &\ge& 1+(1+\rho) \int f_w(y) T(y) dy +\frac{\rho(1+\rho)}{2} \int f_w(y) T^2(y) dy
-\frac{\rho(1+\rho)(1-\rho)}{6} \int f_w(y) T^3(y) dy\\
&=& 1+(1+\rho)(e^{-\theta p}-1) + \frac{\rho(1+\rho)}{2}\left\{ (e^{-\theta p} -1)^2 +
\frac{2p^2}{(1+\rho)^4}+\delta(p^2)\right\}+\rho \delta(p^2)\\
&\ge& 1+(1+\rho)(-\theta p+\frac{\theta^2 p^2}{2}-\frac{\theta^3 p^3}{6})
+ \frac{\rho(1+\rho)}{2}\left\{ (-\theta p+\frac{\theta^2 p^2}{2})^2 +
\frac{2p^2}{(1+\rho)^4}+\delta(p^2)\right\}+\rho \delta(p^2)\\
&=& 1-\frac{\rho}{1+\rho}p+\frac{\rho^3+\rho^2+2\rho}{(1+\rho)^3} \frac{p^2}{2} +\rho \delta(p^2).
\end{eqnarray*}

In other words, we must have
$$
\int \alpha(y) dy = 1-\frac{\rho}{1+\rho}p+\frac{\rho^3+\rho^2+2\rho}{(1+\rho)^3} \frac{p^2}{2} +\rho \delta(p^2).
$$
Thus,
\begin{eqnarray*}
\tilde{r}(p)&=&\sup_{0\le\rho \le 1} -\frac{pz}{\rho} + \frac{{\tilde
    E}_o(p,BPSK,\rho)}{\rho}\\
&=&\sup_{0\le\rho \le 1} -\frac{pz}{\rho} + \frac{-\ln \int \alpha(y) dy}{\rho}\\
&=& \sup_{0\le\rho \le 1} -\frac{pz}{\rho} +
\frac{p}{1+\rho}-\frac{p^2}{(1+\rho)^3} +\delta(p^2).
\end{eqnarray*}

From here, it is easy to check that
$$
\frac{{\tilde E}_o(p,BPSK,\rho)}{p\rho}\rightarrow \frac{1}{1+\rho}
$$
uniformly for $0\le\rho \le 1$ as $p\rightarrow 0.$ Further,
$$
\frac{\frac{{\tilde E}_o(p,BPSK,\rho)}{p\rho}-\frac{1}{1+\rho}}{p}\rightarrow
-\frac{2}{(1+\rho)^3}.
$$
From Theorem~\ref{thm: r dot 0} and Theorem~\ref{thm: r ddot 0}, we know
this implies
\begin{eqnarray*}
\dot{\tilde r}(0)&=&\sup_{0\le\rho\le 1} -\frac{z}{\rho}+\frac{1}{1+\rho}=(1-\sqrt{z})^2;\\
\ddot{\tilde r}(0)&=& \frac{\ddot{{\tilde E}_o}(0,BPSK,\rho^*)}{\rho^*}=-\frac{2}{(1+\rho^*)^3}=-2(1-\sqrt{z})^3.
\end{eqnarray*}

Therefore, BPSK is first-order optimal but not second-order
optimal.

The QPSK calculations are very similar to the BPSK calculations and we can
show that for QPSK
\begin{eqnarray*}
\dot{\tilde r}(0)&=&\sup_{0\le\rho\le 1} -\frac{z}{\rho}+\frac{1}{1+\rho}=(1-\sqrt{z})^2;\\
\ddot{\tilde r}(0)&=& \frac{\ddot{{\tilde E}_o}(0,QPSK,\rho^*)}{\rho^*}=-(1-\sqrt{z})^3,
\end{eqnarray*}
which implies that QPSK is near-optimal.

\section{Proof of Lemma~\ref{lem: sufficient}}\label{sec:
sufficient fading}

It suffices to check (\ref{eq: first condition}) for this choice
of input distributions. When $q_{W_c}$ has i.i.d. entries, we have
the following:
\begin{eqnarray}
E_o(P,q_{W_c},\rho^*,W_c)&\ge& E_o(P,q_{W_c},\rho^*,W_c)|_{\beta=\theta}\nonumber\\
    &=& -\ln E_H\left[\exp\{ -D
    E_o^{NF}(\frac{P|H|^2}{W_c},q,\rho^*)|_{\beta=\theta}\}\right] \label{eq: q1D},
\end{eqnarray}
where $\theta=\frac{\rho^*}{(1+\rho^*)^2}.$ Following
Appendix~\ref{sec: proof all} , we know that if $q$ is symmetric
around $0,$ we have
\begin{equation}
\liminf_{p\rightarrow 0}
\frac{E_o^{NF}(p,q,\rho^*)|_{\beta=\theta}}{p}\ge
\frac{\rho^*}{1+\rho^*}.\label{eq: q2D}
\end{equation}
From Lemma~\ref{lem: upper bound},
$$
\frac{E_o^{NF}(p,q,\rho^*)|_{\beta=\theta}}{p} \le
\frac{\rho^*}{1+\rho^*}.
$$
Thus, actually, if we take $\beta=\theta,$ the limit of
$\frac{E_o^{NF}(p,q,\rho^*)|_{\beta=\theta}}{p}$ exists and is
equal to $\frac{\rho^*}{1+\rho^*}.$

This result also implies
$$
\lim_{W_c\rightarrow \infty} D
E_o^{NF}(\frac{P|H|^2}{W_c},q,\rho^*)|_{\beta=\theta} =
\frac{\rho^* P|H|^2}{1+\rho^*} \quad \textsl{a.e. for $|H|^2\in
R^+.$}
$$

On the other hand, since
$E_o^{NF}(\frac{P|H|^2}{W_c},q,\rho^*)|_{\beta=\theta}\ge 0,$ we
know
$$\exp\{ -D
    E_o^{NF}(\frac{P|H|^2}{W_c},q,\rho^*)|_{\beta=\theta}\}\le 1.$$
Thus, we can apply dominated convergence theorem to (\ref{eq:
q1D}) and we have
\begin{eqnarray*}
\liminf_{W_c\rightarrow \infty} E_o(P,q_{W_c},\rho^*,W_c)&\ge&
\lim_{W_c\rightarrow \infty} -\frac{1}{T_c}\ln E_H\left[\exp\{ -D
E_o^{NF}(\frac{P|H|^2}{W_c},q,\rho^*)|_{\beta=\theta}\}\right]\\
&=&-\frac{1}{T_c}\ln E_H\left[\lim_{W_c\rightarrow \infty} \exp\{
-D
E_o^{NF}(\frac{P|H|^2}{W_c},q,\rho^*)|_{\beta=\theta}\}\right]\\
&=&-\frac{1}{T_c}\ln E_H\left[\exp\{-\frac{\rho^* P|H|^2}{1+\rho^*}\}\right]\\
&=& \frac{1}{T_c}\ln(1+\frac{\rho^* PT_c}{1+\rho^*}).
\end{eqnarray*}

Thus, (\ref{eq: first condition}) holds for this choice of input
distributions. However, there is a little subtlety in applying the
results in AWGN case here, since the $\rho^*$ in AWGN case and the
$\rho^*$ in this paper are different. This can be easily resolved
by observing that the inequality (\ref{eq: q2D}), which we
borrowed from Appendix~\ref{sec: proof all}, is actually true for
any fixed $\rho.$ Thus we can choose $\rho^*$ to be the optimizing
$\rho$ for (\ref{eq: D 00}) and hence the proof.

\bibliographystyle{plain}

\end{document}